\documentclass[letterpaper]{JHEP3}
\usepackage{amssymb}

\usepackage{epsfig}

\usepackage{amsfonts}
\usepackage{amsmath}

 \numberwithin{equation}{section}
 
\usepackage{graphicx}

\newcommand{\insertplot}[5]{\begin{figure}
 \hfill\hbox to 0.05in{\vbox to #5in{\vfill
 \inputplot{#1}{#4}{#5}}\hfill}
 \hfill\vspace{-.1in}
 \caption{#2}\label{#3}
 \end{figure}}
 \newcommand{\inputplot}[3]{
 \special{ps: plotfile #1}
}
\newcounter{fig}   
\newcommand{\vphi}{\varphi}

\newcommand{\beq}{\begin{equation}}
\newcommand{\eeq}{\end{equation}}
\newcommand{\beqs}{\begin{eqnarray}}
\newcommand{\eeqs}{\end{eqnarray}}

\newcommand{\ra}{\rightarrow}
\numberwithin{equation}{section}
\newcommand{\be}{\begin{equation}}
\newcommand{\ee}{\end{equation}}
\newcommand{\bea}{\begin{eqnarray}}
\newcommand{\eea}{\end{eqnarray}}

\usepackage{graphicx}

\abstract{ 
We propose to compute the action and global charges of the asymptotically anti-de Sitter solutions
in Einstein-Gauss-Bonnet theory by adding boundary counterterms
to the gravitational action.
The general expression of the counterterms and the boundary stress tensor is presented for spacetimes
of dimension $d\leq 9$.
We apply this tehnique for several different types of black objects.
Apart from static and rotating black holes, we consider also Einstein-Gauss-Bonnet black string 
solutions with negative cosmological constant.
 
 }
\preprint{ }

\title{Black objects in the Einstein-Gauss-Bonnet theory 
 with negative cosmological constant
 and the boundary counterterm   method}
\author{Yves Brihaye$^{1}$\thanks{E-mail: \texttt{brihaye@umh.ac.be}}~~and\
 Eugen Radu$^2$\thanks{E-mail: \texttt{radu@lmpt.univ-tours.fr}}  \\
$^{1}$Physique-Math\'ematique, Universite de
Mons-Hainaut, Mons, Belgium\\
$^{2}$ Laboratoire de Math\'ematiques et Physique Th\'eorique,
Universit\'e Fran\c{c}ois-Rabelais, Tours, France}

\begin{document}

\section{Introduction}
The AdS/CFT (anti-de Sitter space/conformal field theory)
correspondence \cite{Maldacena:1997re} attracted a lot of attention in the last decade.
This conjecture tells that the partition function in a $d-1$ dimensional CFT
 is given in terms of the classical action in a $d$-dimensional
gravity theory with negative cosmological constant.

In light of this correspondence, asymptotically AdS 
 black objects would offer the possibility of studying some 
aspects of the nonperturbative structure of certain quantum
field theories.  
Therefore it is of interest to find more general gravity solutions 
with negative cosmological constant and to study their physics,
trying to relate it to the physics of the boundary theory.

An interesting case is provided by the  solutions of the
Einstein-Gauss-Bonnet (EGB) theory in $d\geq 5$ dimensions, since
the  Gauss-Bonnet (GB) term appears as the first curvature stringy
correction to general relativity~\cite{1,Myers:1987yn}, when assuming
that the tension of a string is
large as compared to the energy scale of other variables.
The EGB equations contain no higher derivatives of the metric tensor than second order
and it has proven to be free of ghost when expanding around flat space.
In the AdS/CFT correspondence, the introduction of such higher order terms\footnote{ Note, however, that
the first non vanishing corrections that
appear in the type IIB string effective action differ from EGB, involving
eight derivatives, $i.e.$ a term involving four powers of the Riemann tensor together with its
supersymmetric counterparts \cite{1,Grisaru:1986vi}.} 
 corresponds to next to leading order corrections to 
the $1/N$ expansion of the CFT \cite{Fayyazuddin:1998fb}, \cite{Aharony:1998xz}, \cite{Nojiri:2000gv}.

Similar to the case of Einstein gravity,
when computing quantities like the action and mass of EGB solutions,
one encounters infrared divergences, associated with the infinite volume
of the spacetime manifold.
The traditional approach to this problem is to use
a background subtraction whose asymptotic geometry
matches that of the solutions.
However, such a procedure causes the resulting physical
quantities to depend on the choice of reference background; furthermore, it is not possible in
general to embed the boundary surface into a background spacetime. 
For asymptotically AdS
solutions of the Einstein gravity, one can instead deal with these divergences via 
the counterterm method inspired
by the AdS/CFT correspondence \cite{Balasubramanian:1999re}.
The procedure consists of adding to the action suitable 
boundary counterterms, which are built up with
curvature invariants of  the boundary metric 
and thus obviously they do not alter the bulk equations of motion.
This yields a well-defined Brown-York boundary stress tensor  \cite{Brown:1993br} and
a finite action and mass of the system\footnote{This approach has been generalized for spacetimes 
which are not asymptotically AdS  and the cosmological constant is replaced by a dilaton scalar potential, see $e.g.$ \cite{Cai:1999xg}}.

In principle, there are no obstacles in computing the action and global charges of EGB solutions 
by using a similar approach.
At any given dimension one can write down only a finite number of counterterms that do not vanish at infinity
and this does not depend upon the bulk theory is Einstein or GB.
However, the presence in this case of a new length scale implies a complicated
expression for the coefficients of the boundary  counterterms and makes the procedure
technically much more involved.
To our knowledge,  the only
cases considered in the literature\footnote{An alternative regularization prescription for 
any Lovelock theory with AdS asymptotics has been proposed in \cite{kofinas}, \cite{Miskovic:2007mg}, \cite{Kofinas:2007ns}. 
This approch uses boundary terms with
explicit dependence on the extrinsic curvature $K_{ab}$,
also known as Kounterterms \cite{Olea:2006vd}.} 
correspond  to $d=5$ solutions \cite{Cvetic:2001bk}, \cite{Nojiri:2001ae},
\cite{Brihaye:2008kh}, and solutions in arbitrary dimension 
possesing a zero curvature\footnote{In this case all curvature invariants are zero except for a constant.
Therefore the only possible boundary counterterm is proportional to the volume of the boundary,
regardless of the number of dimensions.} 
boundary \cite{Dehghani:2006dh}.

The first aim of this paper is to generalize the 
boundary counterterms and the quasilocal stress energy
tensor in \cite{Balasubramanian:1999re} to EGB theory.
Our results are valid for configurations with $d\leq 9$, although
a general counterterm expression is also conjectured.
Upon application of the Gibbs-Duhem relation to the partition function, this 
yields an expression for the entropy of a black objects which contains the 
GB corrections.

In the second part of this paper we apply this general formalism to  
asymptotically AdS black holes and black strings in EGB theory.
In the static black hole case, where a closed form solution is known, the expressions we find for the mass-energy and entropy 
 agree with known results in the literature.
Apart from these static solutions, we consider also rotating black holes  with equal magnitude angular
momenta in a odd number of spacetime dimensions.
The second type of black objects studied in this work are the EGB generalization of the
asymptotically AdS black strings considered in \cite{Copsey:2006br},
\cite{ Mann:2006yi}.
We find that these solutions present some new qualitative features
as compared to the case of Einstein gravity.
 
  Our  paper is structured as follows: in the next section we explain 
the model and describe 
the computation of the physical quantities of the solutions such as 
their action and  mass-energy. 
Our proposal for the general counterterm expression in EGB theory 
is also presented there. 
The general 
properties of the black hole solutions are presented in Section $3$,
while in Section $4$ we present 
the new results obtained in the case of  EGB
black strings 
Their construction is based on both analytical and numerical techniques.
 We give our conclusions and 
remarks in the final Section.

\section{The formalism}
\subsection{The action and field equations}
 We consider the EGB action with a negative cosmological constant $\Lambda=-(d-2)(d-1)/2\ell^2$ 
\begin{eqnarray}
\label{action}
I=\frac{1}{16 \pi G}\int_\mathcal{M}~d^dx \sqrt{-g} \left(R-2 \Lambda+\frac{\alpha}{4}L_{GB} \right),
\end{eqnarray}
where $R$ is the Ricci scalar  and
\begin{eqnarray}
\label{LGB}
L_{GB} =R^2-4R_{\mu \nu}R^{\mu \nu}+R_{\mu \nu \sigma \tau}R^{\mu \nu \sigma \tau},
\end{eqnarray}
is the GB term.
In four dimensions this is a topological invariant; in higher dimensions it is 
the most general quadratic expression which preserves the property that the equations of motion involve 
only second order derivatives of the metric.
$L_{GB}$ can also be viewed as the second order term in the Lovelock theory of gravity constructed 
from vielbein, the spin connection and their exterior derivatives without using the Hodge dual,
such that the field equations are second order  \cite{Lovelock:1971yv}, \cite{Mardones:1990qc}.
The constant $\alpha$ in (\ref{action}) is the GB coefficient with dimension $(length)^2$ and is positive
in the string theory.
We shall therefore restrict in this work to the case $\alpha>0$, although the counterterm expression does not depend on this choice. 

The variation of the action (\ref{action}) with respect to the metric
tensor results in the equations of the model
\begin{eqnarray}
\label{eqs}
R_{\mu \nu } -\frac{1}{2}Rg_{\mu \nu}+\Lambda g_{\mu \nu }+\frac{\alpha}{4}H_{\mu \nu}=0~,
\end{eqnarray}
where
\begin{equation}
\label{Hmn}
H_{\mu \nu}=2(R_{\mu \sigma \kappa \tau }R_{\nu }^{\phantom{\nu}%
\sigma \kappa \tau }-2R_{\mu \rho \nu \sigma }R^{\rho \sigma }-2R_{\mu
\sigma }R_{\phantom{\sigma}\nu }^{\sigma }+RR_{\mu \nu })-\frac{1}{2}%
L_{GB}g_{\mu \nu }  ~.
\end{equation}
For a well-defined variational principle, one has to supplement the 
action (\ref{action}) with the Gibbons-Hawking surface term \cite{GibbonsHawking1}
\begin{equation}
I_{b}^{(E)}=-\frac{1}{8\pi G}\int_{\partial \mathcal{M}}d^{d-1}x\sqrt{-\gamma }K~,
\label{Ib1}
\end{equation}
and its counterpart for the GB gravity  \cite{Myers:1987yn} 
\begin{equation}
I_{b}^{(GB)}=-\frac{\alpha}{16\pi G}\int_{\partial \mathcal{M}}d^{d-1}x\sqrt{-\gamma }%
 \left( J-2{\rm G}_{ab} K^{ab}\right)  ~,
\label{Ib2}
\end{equation}
where $\gamma _{ab }$ is the induced metric on the boundary,  $K$ is  the
trace of the extrinsic curvature of the boundary,
  ${\rm G}_{ab}$ is the Einstein tensor of the metric $\gamma _{ab}$ and $J$ is the
trace of the tensor
\begin{equation}
J_{ab}=\frac{1}{3}%
(2KK_{ac}K_{b}^{c}+K_{cd}K^{cd}K_{ab}-2K_{ac}K^{cd}K_{db}-K^{2}K_{ab})~.
\label{Jab}
\end{equation}
We shall consider spacetimes of negative constant curvature in the asymptotic region, 
which implies the asymptotic
expression of the Riemann tensor
$R_{\mu \nu}^{~~\lambda \sigma}=-(\delta_\mu^\lambda \delta_\nu^\sigma
-\delta_\mu^\sigma \delta_\nu^\lambda)/\ell_c^2$,
where $\ell_c$ is the new effective radius of the AdS space in EGB theory\footnote{For a precise definition of asymptotically AdS spacetime in higher curvature gravitational theories, see $e.g. $ \cite{Okuyama:2005fg}.}.
We have found  convenient to write
~
\begin{eqnarray}
\label{lc}
\ell_c=\ell\sqrt{\frac{1+U}{2}},~~{\rm~~with~~~~}U=\sqrt{1-\frac{\alpha(d-3)(d-4)}{\ell^2}},
\end{eqnarray}
which results in a compact form for the couterterms and
a simpler expression of the black string asymptotics.
One should also notice the existence of an upper bound for the  GB coefficient,
$\alpha\leq \alpha_{max}=\ell^2/(d-3)(d-4)$, which holds for all asymptotically AdS solutions.

\subsection{The counterterms and boundary stress tensor}
The action and global charges of the EGB-$\Lambda$ solutions are computed by using
a suitable generalization of the procedure proposed by Balasubramanian and Kraus 
\cite{Balasubramanian:1999re}  for Einstein gravity
with negative cosmological constant. This technique was inspired by the 
AdS/CFT correspondence (since quantum field theories in general contains 
counterterms)
and consists of adding to the action suitable 
boundary counterterms $I_{ct}$, which are functionals only of 
curvature invariants of the induced metric on the boundary. 
The number of terms that appears grows with the dimension of the spacetime.

To regularize the action of the $d<8$  solutions, we supplement the general action (which contains the surface terms for
Einstein and GB gravity)  with  the following boundary counterterms
\begin{eqnarray}
\label{Lagrangianct} 
I_{\mathrm{ct}}^0 &=&\frac{1}{8\pi G}\int_{\partial \mathcal{M}} d^{d-1}x\sqrt{-\gamma 
}
\bigg\{
  -(\frac{d-2}{\ell_c })(\frac{2+U}{3})
 -\frac{\ell_c \mathsf{\Theta } \left( d-4  \right) } {2(d-3)}(2-U)\mathsf{R}
 \\
 \nonumber
 &&-\frac{\ell_c ^{3}\mathsf{\Theta }\left( d-6\right) }{2(d-3)^{2}(d-5)}
 \left[ U\bigg( \mathsf{R}_{ab}\mathsf{R}^{ab}-\frac{d-1}{4(d-2)}\mathsf{R}^{2} \bigg)
 -\frac{d-3}{2(d-4)}(U-1)L_{GB} 
  \right] 
\bigg\},
\end{eqnarray}
where $\mathsf{R}$, $\mathsf{R}^{ab}$  and $L_{GB}$ are the curvature, the 
Ricci tensor and the GB term associated with the induced metric $\gamma $. 
Also, $\mathsf{\Theta}(x)$ is the step-function with  $\mathsf{\Theta 
}\left( x\right) =1$ provided $x\geq 0$, and zero otherwise.
One can see that as $\alpha \to 0$ ($U \to 1$), one recovers the known counterterm
expression in the Einstein gravity \cite{Balasubramanian:1999re}, \cite{Emparan:1999pm}, \cite{Mann:1999pc}.

A similar expression can be written for higher dimension than seven, with new terms entering at any even $d$.
However, their complexity strongly increases with $d$.
Based on the relations we have derived for $d<10$, we conjecture the following general 
expression for the boundary counterterm in the EGB theory\footnote{Following 
 \cite{Balasubramanian:1999re}, \cite{Emparan:1999pm}, the $d\leq 9$ counterterms
were obtained by demanding cancellation of divergencies for a number of solutions in EGB theory.
For example, for $d\leq 7$, the only geometric terms that possible do not vanish at infinity
are $1/\ell_c$, $\mathsf{R}$ and $\mathsf{R}^2$, $\mathsf{R}_{ab}\mathsf{R}^{ab}$, 
$\mathsf{R}_{abcd}\mathsf{R}^{abcd}$. However, the last term does not appear in the Einstein gravity
expression. Therefore, we have found convenient to express the $d\leq 7$ counterterm (\ref{Lagrangianct}) 
as the sum of a Einstein-gravity part (multiplied with some $U$-dependent factors) and a
GB term with a factor of $U-1$ in front of it ($i.e.$ it vanishes in the Einstein gravity limit of the bulk theory).
The same approach yields a relatively simple expression of the counterterm also for $d=8,9$.
For the generality of the results proposed here, see the comments in the last Section of this work.
}
\begin{eqnarray}
\label{Ict-gen}
I_{\mathrm{ct}}^0 &=&\frac{1}{8\pi G}\int_{\partial \mathcal{M}} d^{d-1}x\sqrt{-\gamma }
\bigg\{  
\sum_{k\geq 1} \mathsf{\Theta } \left( d-2k  \right)\bigg(f_1(U)L_E+f_2(U){\cal L}_{(k-1)} 
\bigg)
\bigg\} ,
\end{eqnarray}
where $L_E$ is the corresponding $k$-th part of
the counterterm lagrangean for a theory with only Einstein gravity in the bulk 
(with the length scale $\ell$ in front of it replaced
by the new effective AdS radius $\ell_c$) and ${\cal L}_{(k-1)}$ is the $(k-1)$ term in the Lovelock hierarchy.
The functions $f_1(U)$, $f_2(U)$ have the general expression 
\begin{eqnarray}
f_1=1+c_1(U-1),~~~f_2= \ell_c^{2k-3}c_2(U-1),
 \end{eqnarray}
where $c_1,c_2$ are $d$-dependent coefficients. 
The series 
truncates for any fixed dimension, with new terms entering at every 
new even value of $d$.

The relation (\ref{Lagrangianct}) contains
already the cases $k=1,2,3$. For $k=1$ one finds $c_1=0$, $c_2=-(d-2)/3$, while $L_E=-(d-2)/\ell_c$, ${\cal L}_{(0)}=1$.
Taking $k=2$ yields $c_1=0$, $c_2=1/(2(d-3))$ and $L_E=- \ell_c \mathsf{R}/2(d-3)$, ${\cal L}_{(1)}=\mathsf{R}$.
The value $k=3$ implies the following expression for the new terms in (\ref{Ict-gen})
\begin{eqnarray}
\label{k=3}
&&c_1=1, ~~c_2=\frac{1}{4(d-3)(d-4)(d-5)}~,
\\
\nonumber
&&L_E=-\frac{\ell_c ^{3}}{2(d-3)^{2}(d-5)} \left( \mathsf{R}_{ab}\mathsf{R}^{ab}-\frac{d-1}{4(d-2)}\mathsf{R}^{2} \right),~~
{\cal L}_{(2)}=L_{GB}.
 \end{eqnarray}
 The case $k=4$ is more involved, with
 \begin{eqnarray} 
 &&L_E=\frac{\ell_c ^{5}}{(d-3)^{3}(d-5)(d-7)}\bigg( \frac{3d-1}{%
4(d-2)}\mathsf{R} \mathsf{R}^{ab}R_{ab}-\frac{\left( d-1\right) (d+1)}{16(d-2)^{2}}\mathsf{R}^{3}  
\\
\nonumber
 &&{~~~~~~~~}
 -  2\mathsf{R}^{ab}\mathsf{R}^{cd}\mathsf{R}_{acbd} +
  \frac{d-3}{2(d-2)}\mathsf{R}^{ab}\nabla _{a}\nabla_{b}\mathsf{R}
  -\mathsf{R}^{ab}\nabla ^{2}\mathsf{R}_{ab}
  +\frac{1}{2(d-2)}\mathsf{R}\nabla ^{2}\mathsf{R}
\bigg),
 \end{eqnarray}
 as given in \cite{Das:2000cu}, and the third Lovelock term
  \begin{eqnarray} 
  &&{\cal L}_{(3)} =2\mathsf{R}^{abcd }\mathsf{R}_{cd  ef }
 \mathsf{R}_{
\phantom{ef }{ab }}^{ef }
+8\mathsf{R}_{~~cd}^{ab }\mathsf{R}_{~~bf}^{ ce }\mathsf{R}_{~~ae}^{df}
+24\mathsf{R}^{abcd}\mathsf{R}_{cdbe}\mathsf{R}_{a}^{e} 
 \\
 &&{~~~~~~~~}+3\mathsf{R}\mathsf{R}^{abcd }\mathsf{R}_{cdab }
+24\mathsf{R}^{abcd }\mathsf{R}_{ ca}\mathsf{R}_{db }
+16\mathsf{R}^{ab }\mathsf{R}_{bc}\mathsf{R}_{a}^{c}
-12\mathsf{R}\mathsf{R}^{ab }\mathsf{R}_{ab }
+\mathsf{R}^{3}~,
\notag
 \end{eqnarray}
the constants $c_1,c_2$ in the expression of $f_1,f_2$ being 
\begin{eqnarray}
\label{k=4}
&&c_1=31/30,~~c_2=-19/57600,~~~{\rm for~}d=8,
\\
\nonumber
&&c_1=2365/2313,~~c_2=-149/2664576,~~~{\rm for~}d=9.
 \end{eqnarray}
Varying the total action with respect to the boundary metric $\gamma _{ab}$
\begin{equation}
T_{ab}=\frac{2}{\sqrt{-\gamma }}\frac{\delta }{\delta \gamma ^{ab}}\left(
I+I_{b}^{(E)}+I_{b}^{(GB)}+I_{\text{ct}}^0 \right)  
\label{Tab}
\end{equation}
results in the boundary stress-energy tensor:
\begin{eqnarray}
 T_{ab}&=&
K_{ab}-\gamma _{ab}K
 +\frac{{\alpha}}{2} (Q_{ab}-\frac{1}{3}Q\gamma_{ab}) 
-\frac{d-2}{\ell_c}\gamma _{ab}(\frac{2+U}{3})
+\frac{\ell_c
\mathsf{\Theta } \left( d-4  \right) }{d-3}(2-U)%
\left( \mathsf{R}_{ab}-\frac{1}{2}\gamma _{ab}\mathsf{R} \right)  
\notag 
\\
&&+\ell_c^{3}\mathsf{\Theta } \left( d-6  \right) 
\bigg\{
\frac{U}{(d-3)^{2}(d-5)}
 \bigg(
   -\frac{1}{2}\gamma _{ab}\left(
\mathsf{R}_{cd}\mathsf{R}^{cd}-\frac{(d-1)}{4(d-2)}\mathsf{R}^{2}\right)  
\notag 
\\
\label{TabCFT}
&&  
-\frac{(d-1)}{2(d-2)}%
\mathsf{RR}_{ab} 
+2\mathsf{R}^{cd}\mathsf{R}_{cadb}-\frac{d-3}{2(d-2)}\nabla _{a}\nabla _{b}\mathsf{R}+\nabla
^{2}\mathsf{R}_{ab}-\frac{1}{2(d-2)}\gamma _{ab}\nabla ^{2}\mathsf{R } 
\bigg )
\\
\nonumber
&&-\frac{ U-1 }{2(d-3) (d-4)(d-5)}
\mathsf{H}_{ab}
\bigg \}+\dots ,
\end{eqnarray}
where \cite{Davis:2002gn}
\begin{eqnarray}
&Q_{ab}= 
2KK_{ac}K^c_b-2 K_{ac}K^{cd}K_{db}+K_{ab}(K_{cd}K^{cd}-K^2)
+2K \mathsf{R}_{ab}+\mathsf{R}K_{ab}
-2K^{cd}\mathsf{ R}_{cadb}-4 \mathsf{R}_{ac}K^c_b,~{~~~~~~}
\end{eqnarray}
and $\mathsf{H}_{ab}$ given by (\ref{Hmn}) in terms of the boundary metric $\gamma_{ab}$.

The expression (\ref{TabCFT}) is valid for  $ d\leq 7$. 
For $d\geq 8$, the boundary stress tensor receives a supplementary constribution from the 
$k=4$ term in (\ref{Ict-gen}).
The Einstein part of it can be found in Ref. \cite{Das:2000cu};
the other contribution represents the third  Lovelock tensor as derived in Ref. \cite{MuellerHoissen:1985mm}
(from (\ref{Ict-gen}), (\ref{Tab}) they are both multiplied with $U$-dependent factors).
These extra-terms  are very long and we prefere do not include them here.

However, for odd values of $d$, 
the counterterms proposed above may fail to regularize the action. 
This is the case of the black string solutions we shall discuss in 
in the Section 4, or of the EGB Euclidean AdS$_d$ metric
$ ds^2=\frac{dr^2}{1+r^2/\ell_c^2}+r^2d\Omega_{d-1}^2$
 (where $d\Omega^2$ is the unit metric on the sphere).
Already in the $\alpha=0$ case, the action of these solutions presents a logarithmic
divergence, whose coefficient is related to the conformal
Weyl anomaly in the dual theory defined in a even dimensional spacetime.

However, the  
divergences are removed by adding the following extra term to  (\ref{action})  
\begin{eqnarray}
I_{\mathrm{ct}}^{s} &=&\frac{1}{8\pi G_d}\int_{\partial \mathcal{M}} d^{d-1}x\sqrt{-\gamma 
}\log(\frac{r}{\ell_c})\bigg\{  
\mathsf{\delta }_{d,5}\frac{\ell_c^3 }{8}
\bigg[
U(\frac{1}{3}\mathsf{R}^2-\mathsf{R}_{ab}\mathsf{R}^{ab})
-(U-1)(\mathsf{R}^2
\\
\nonumber  
&&-4\mathsf{R}_{ab}\mathsf{R}^{ab}+\mathsf{R}_{abcd}\mathsf{R}^{abcd}
\bigg] 
-\delta_{d,7}
\frac{\ell_c^{5}}{128}\bigg[
 \frac{1}{18}(19U-1)
\bigg(
\mathsf{RR}^{ab}\mathsf{R}_{ab}
-\frac{3}{25}\mathsf{R}^{3} 
-2\mathsf{R}^{ab}\mathsf{R}^{cd}\mathsf{R}_{acbd}
\\
\nonumber  
&&
-\frac{1}{10}\mathsf{R}^{ab}\nabla _{a}\nabla 
_{b}\mathsf{R}+\mathsf{R}^{ab}\Box \mathsf{R}_{ab}
-\frac{1}{10}\mathsf{R}\Box 
\mathsf{R}
\bigg) 
-\frac{11}{54}(U-1){\cal L}_{(2)}
\bigg] +\dots
\bigg\}~,
\end{eqnarray}%
 (with $r$ the radial coordinate normal to the boundary).
This gives a supplementary contribution to the boundary stress 
tensor which removes the logaritmic divergencies from the
solutions' global charges.
Note that a logarithmic contribution to the counterterms also appears naturally in a
 formulation of the holographic renormalization procedure in terms of the extrinsic
curvature \cite{Papadimitriou:2004ap}.

The computation of the global charges associated with the Killing symmetries of
the boundary metric is done in a similar way to the case of Einstein gravity.
 One assumes that the boundary submanifold can be foliated in a standard ADM  form
\beqs
\gamma_{ab}dx^adx^b=-N^2dt^2+\sigma_{ij}(dy^i+N^idt)(dy^j+N^jdt),
\label{ADMboundary}
\eeqs
where $N$ and $N^i$ are the lapse function, respectively 
the shift vector, and $y^i$, $i=1,\dots,d-2$ are the intrinsic 
coordinates on a closed surface $\Sigma$ of constant time 
$t$ on the boundary.  
Then a conserved charge 
\begin{equation}
{\frak Q}_{\xi }=\oint_{\Sigma }d^{d-2}y\sqrt{\sigma}u^{a}\xi ^{b}T_{ab},
\label{Mcons}
\end{equation}%
can be associated with the closed surface $\Sigma $ 
(with normal $u^{a}$), provided the boundary geometry 
has an isometry generated by a Killing vector $\xi ^{a}$. 
For example,  the conserved mass/energy $M$ is the charge associated 
with the time translation symmetry, with $\xi =\partial /\partial t$. 

The background metric upon which the dual field theory resides is 
$h_{ab}=\lim_{r \rightarrow \infty} \frac{\ell^2_c}{r^2}\gamma_{ab}$.
The expectation value of the stress tensor of the dual theory 
can be computed using the  relation 
\cite{Myers:1999ps}:
\begin{eqnarray} 
\label{Tik-CFT}
\sqrt{-h}h^{ab}<\tau _{bc}>=\lim_{r\rightarrow \infty }\sqrt{-\gamma 
}\gamma
^{ab}T_{bc},
\end{eqnarray}
and is expected to present a nontrivial dependence on the parameter $\alpha$.

The Hawking 
temperature $T_H$ of the black objects is found by demanding regularity of the 
Euclideanized manifold, or equivalently,  by evaluating the surface gravity. 
The entropy of the black hole objects is computed in this work using the (Euclidean) path-integral formalism\footnote{Note,
however, that not all  solutions with
Lorentzian signature present reasonable Euclidean counterparts (see, $e.g.$ \cite{Astefanesei:2005ad}), in which case
one is forced again to consider a 'quasi-Euclidean' approach as described in \cite{quasi}.
In this case the action $I$ is regarded as a functional 
over complex metrics that are obtained from the real, 
stationary, Lorentzian metrics by using a transformation 
that mimics the effect of the Wick rotation $t\ra i\tau$. 
The values of 
the extensive variables of the complex 
  metric that extremize the path integral are the same as the values 
 of these variables corresponding to the initial Lorentzian metric.}. 
In this approach, the gravitational 
thermodynamics is based on the general relation \cite{Hawking:ig} 
\[
Z=\int D\left[ g\right] D\left[ \Psi \right] e^{-I\left[ g,\Psi 
\right]
}\simeq e^{-I_{cl}}, 
\]%
where $D\left[g\right]$ is a measure on the space of metrics $g$, $D\left[\Psi\right]$ a measure on the 
space of matter fields $\Psi$
and $I_{cl}$ is the classical action evaluated on the equations of motion of the gravity/matter system.
This yields an expression for the entropy (with $\beta=1/T_H$)
\begin{equation}
S=\beta (M-\mu _{i}{\frak C}_{i})-I_{cl},  
\label{GibbsDuhem}
\end{equation}%
upon application of the Gibbs-Duhem relation to the partition 
function, with chemical potentials ${\frak C}_{i}$ and
conserved charges $\mu _{i}$  (see $e.g.$ \cite{Mann:2003 Found}).
In Einstein gravity, the entropy computed in this way is one quarter of the event horizon area for any black object.
However,  a GB term in the action may provide a nonzero contribution to $S$.
Finding this correction within the Euclidean approach requires a consideration of
 concrete configurations, no general result being available.
However, all solutions should satisfy the
 first law of thermodynamics  
\begin{equation}
dS=\beta (dM-\mu _{i}d{\frak C}_{i}). 
\label{1stlaw}
\end{equation} 

\section{EGB-AdS black holes }

\subsection{Static solutions}
Due to the nonlinearity of the field equations, it is very
difficult to find  exact  solutions of the EGB equations.
However, the static spherical and topological black holes are known in closed
form, as well as their generalization with an extra U(1) field.

The counterparts of the Schwarzschild solution in EGB theory with 
negative cosmological constant were found in  \cite{Deser}, \cite{Cai:2001dz}
and have a line element
\begin{eqnarray}
\label{SGB}
ds^2=\frac{dr^2}{f(r)}+r^2 d\Sigma^2_{k,d-2}-f(r) dt^2
\end{eqnarray}%
where the $(d{-}2)$--dimensional metric $d\Sigma^2_{k,d-2}$ is
\begin{equation}
d\Sigma^2_{k,d-2} =\left\{ \begin{array}{ll}
\vphantom{\sum_{i=1}^{d-2}}
 d\Omega^2_{d-2}& {\rm for}\; k = +1\\
\sum_{i=1}^{d-2} dx_i^2&{\rm for}\; k = 0 \\
\vphantom{\sum_{i=1}^{d-3}}
 d\Xi^2_{d-2} &{\rm for}\; k = -1\ ,
\end{array} \right.
\end{equation}
and $d\Omega^2_{d-2}$ denoting the unit metric on $S^{d-2}$. By $H^{d-2}$ 
we will understand the $(d{-}2)$--dimensional hyperbolic space, whose unit 
metric  $d\Xi^2_{d-2}$ can be obtained by analytic continuation of 
that on $S^{d-2}$.

The function $f(r)$ presents a complicated dependence on $\ell,\alpha$
(here we restrict to the branch of solutions which are well behaved in the $\alpha \to 0$ limit)
\begin{eqnarray}
\label{f-BH}
 f(r) = k+\frac{2r^2}{\alpha (d-3)(d-4)}
 \bigg (
 1-\sqrt{1+\alpha (d-3)(d-4)(\frac{\mu}{r^{d-1}}-\frac{1}{\ell^2})}
 \bigg),
\end{eqnarray}%
with $\mu$ a constant. 

The Hawking temperature of the black holes is $T_H=f'(r_h)/(4 \pi)$, where a prime denotes 
the derivative with respect the radial coordinate and
 $r_h$ is the largest positive root of $f(r)$, typically associated to the outer horizon of a black hole.

One can compute
the action of these solutions  in a simple way by noticing the relation
\begin{eqnarray}
&\frac{1}{2}\left( R-2 \Lambda +\frac{1}{4}\alpha L_{GB}^{(1)}\right)= r^{2-d}
\left(-\frac{1}{2}r^{d-2}f'+\frac{1}{4}\alpha(d-2)(d-3)r^{d-4}f'(f-k)\right)'~{}.
\end{eqnarray}%
A straightforward computation\footnote{The supplementary counterterm $I_{ct}^s$
vanishes for the boundary geometry of the black hole solutions.}  shows that the contribution of the bulk action
at infinity together with the boundary terms $I_{b}^{(E)}+I_{b}^{(GB)}+I_{\text{ct}}^0$
is equal to $\beta M$, where  $M$ is the
 mass-energy of the solutions as computed from the boundary stress tensor
\begin{eqnarray}
M=\frac{(d-2)V_{k,d-2}}{16 \pi G}\mu+M_k^0~.
\end{eqnarray}%
In the above expression
\begin{eqnarray}
M_k^0=\frac{V_{k,d-2}}{16 \pi G}
\left(
\frac{3k^2}{4} \ell_c^2(3U-2) \delta_{d,5}
-\frac{5k }{8} \ell_c^4\frac{5U-2}{3} \delta_{d,7}
+\frac{35k^2}{64} \ell_c^6\frac{7U-2}{5} \delta_{d,9}+\dots
\right )
\end{eqnarray}%
is the Casimir mass-energy\footnote{Note that the expression of the Casimir energy agrees with that found in  \cite{kofinas} 
by using Kounterterms regularisation.}
and $V_{k,d-2}$ is the (dimensionless) volume associated with the   metric $d\Sigma^2_{k,d-2}$.
Following \cite{Emparan:1999pm}, 
one can extrapolate the Casimir term to
\begin{eqnarray}
\label{Casimir}
M_k^0={V_{k,d-2}\over 8\pi G}(-k)^{(d-1)/2}{(d-2)!!^2\over (d-1)!} 
\left(\frac{(d-2)U-2}{d-4} \right)\ell_c^{d-3} 
.
\end{eqnarray}%
From (\ref{GibbsDuhem}) one finds the EGB black hole entropy 
\begin{eqnarray}
\label{BH-entropy}
 S=S_0+S_{c}~~~{\rm with}~~~S_0=\frac{A_H}{4G}~,
~~~~S_c=\alpha  \frac{V_{k,d-2}}{4G}\frac{k}{2} r_h^{d-4}(d-2)(d-3),
\end{eqnarray}  
with  the event horizon area $A_H=r_h^{d-2}V_{k,d-2}$.
One can easily verify that the first law of thermodynamics (\ref{1stlaw}) also holds, with $\mu _{i}={\frak C}_{i}=0$.

The expressions of mass (without the Casimir term) and entropy agree with previous results in the literature
found by using a different approach  \cite{Cho:2002hq}, \cite{Dutta:2006vs}, \cite{Clunan:2004tb}, \cite{Padilla:2003qi}.
The Noetherian (or Wald's) approach \cite{Wald:1993nt} is particularly interesting, presenting an expression of $S$ which is
holographic in spirit  \cite{Dutta:2006vs}.
As discussed in   \cite{Clunan:2004tb}, the  entropy of the black hole solutions (\ref{SGB}) can be written
as a integral over the event horizon
\begin{eqnarray}
\label{S-Noether} 
S=\frac{1}{4G}\int_{\Sigma_h} d^{d-2}x \sqrt{\tilde h}(1+\frac{\alpha}{2}\tilde R),
\end{eqnarray} 
(where $\tilde h$ is the determinant of the induced metric on the horizon and $\tilde R$ is the event horizon curvature),
which agrees with the result (\ref{BH-entropy}).
This relation appears to be universal, being satisfied by the other EGB black objects discussed in this work.

The stress tensor of the dual theory 
computed according to (\ref{Tik-CFT}) has the same expression as in the $\alpha=0$ case
\begin{eqnarray}
 8 \pi G <\tau _{a}^b>=\frac{M}{2\ell_c^{d-2}}(\delta_a^b+(d-1)u_a u^b)~, 
\end{eqnarray}
where $u_a=\delta_a^t$.
This tensor is finite, covariantly conserved and manifestly traceless and presents a nontrivial dependence of the GB coefficient $\alpha$.

The GB term gives rise to some interesting effects on the thermodynamics of black holes
in AdS space.
First, as observed in \cite{Cvetic:2001bk}, \cite{Nojiri:2002qn}, 
for $k=-1$ topological black holes, the second term in (\ref{BH-entropy}) can make the whole
expression negative for sufficiently small black holes.
In the $k=0$ case, where the horizon is flat, the GB term has no effect on the expression for entropy,
which is simply the area of the horizon.
For spherically symmetric solutions, a
locally stable small black holes branch appears for $d=5$, which
is absent in the case without the GB term. However, for $d\geq 6$, the thermodynamic
behavior of the EGB black holes is qualitatively similar to the case with $\alpha=0$. 
Detailed discussions of the thermodynamics of EGB black holes
can be found in  \cite{Cvetic:2001bk}, \cite{Nojiri:2001ae}, \cite{Cho:2002hq}.

\subsection{Rotating black holes with equal magnitude angular momenta}
The computation of the global charges and entropy of the
rotating black holes in EGB theory represents another nontrivial aplication
of the general formalism in Section 2.
No exact solutions are available in this case\footnote{Slowly rotating EGB 
solutions have been considered recently in \cite{Kim:2007iw} within a perturbative approach.}, except for the $k=0$ configurations in \cite{Dehghani:2002wn}.
However, these exact solutions  are essentially
obtained by boosting the static configurations with flat horizon; thus locally they are equivalently to the static ones but not globally.

In this subsection we consider a class of EGB black holes in a odd number of spacetime 
dimensions $d=2N+1$ ($N\geq 2$), possessing equal magnitude angular momenta and a spherical horizon topology.
This factorizes the 
angular dependence \cite{Kunz:2006eh} and reduces the problem 
to studying the solutions of four differential equations with dependence 
only on the radial variable $r$. 
The explicit computation of 
the black holes' action and boundary stress tensor also simplifies drastically in this case.

These
solutions are found for the following parametrization of the metric\footnote{For $\alpha=0$ ($i.e.$ no GB term), 
these solutions are a subset of the general configurations with the maximal 
number of rotation parameters
discussed in \cite{Gibbons:2004js}, with
$
f(r)=1
+ {r^2}/{\ell^2}
- {2{\hat M}\Xi}/{r^{d-3}}
+ {2{\hat M}{\hat a}^2}/{r^{d-1}},$ $
h(r)=r^2\left(1+ {2{\hat M}{\hat a}^2}/{r^{d-1}}\right),$ $
w(r)= {2{\hat M}{\hat a}}/{r^{d-3}h(r)},$ $
g(r)=r^2,$ $ b(r)= {r^2f(r)}/{h(r)},
$
where ${\hat M}$ and ${\hat a}$ are two constants related to the solution's mass and 
angular momentum and $\Xi=1-{\hat a}^2/\ell^2$.}
\begin{eqnarray}
& ds^2 = -b(r)dt^2  +  \frac{ dr^2}{f(r)} + 
g(r)\sum_{i=1}^{N-1}
  \left(\prod_{j=0}^{i-1} \cos^2\theta_j \right) d\theta_i^2  
\nonumber
 \\  
&+h(r) \sum_{k=1}^N \left( \prod_{l=0}^{k-1} \cos^2 \theta_l
  \right) \sin^2\theta_k \left( d\vphi_k - w(r)
  dt\right)^2 
  \label{metric-rot}
 \\ 
& +(g(r)-h(r)) \left\{ \sum_{k=1}^N \left( \prod_{l=0}^{k-1} \cos^2
  \theta_l \right) \sin^2\theta_k  d\vphi_k^2 \right. 
  -\left. \left[\sum_{k=1}^N \left( \prod_{l=0}^{k-1} \cos^2
  \theta_l \right) \sin^2\theta_k   d\vphi_k\right]^2 \right\},
\nonumber
\end{eqnarray}
where $\theta_0 \equiv 0$, $\theta_i \in [0,\pi/2]$ 
for $i=1,\dots , N-1$, $\theta_N \equiv \pi/2$, 
$\vphi_k \in [0,2\pi]$ for $k=1,\dots , N$.
The  metric gauge choice we consider here 
is $h(r)=r^2$. 
Note also that the $k=1$ static black holes (\ref{SGB}) are recovered for $w(r)=0$, $ h(r)=g(r)=r^2$ and $b=f$ as given by (\ref{f-BH}).

Although only the $d=5$ case has been studied so far in the literature by using numerical methods \cite{Brihaye:2008kh}, similar
solutions should exist for all $N$.
A discussion of the properties of these stationary black holes is beyond the purposes of this work.
Here we shall present only a computation of their mass-energy, angular momentum and entropy,  in which case only the asymptotic
form of the metric is required, looking for the
effects at this level of the GB term\footnote{For a different computation of
mass and angular momentum of rotating black holes in EGB gravity, see \cite{Deruelle:2004bj}, \cite{tekin}.}.

By solving the EGB field equations (\ref{eqs})  for large $r$, one find that these rotating black holes 
present the following asymptotics in terms of the arbitrary constants $\bar f,\bar b,\bar j$
 \begin{eqnarray}
\label{inf1}
&&f= 1+ \frac{r^2}{\ell_c^2}  
+\bar f \left( \frac{\ell_c }{r } \right)^{d-3}+\dots,~~~
b= 1+ \frac{r^2}{\ell_c^2}  + \bar b \left( \frac{\ell_c }{r } \right)^{d-3}+\dots,~~~
\\
\nonumber
&&h/r^2= 1+ (\bar f-\bar b) \left( \frac{\ell_c }{r } \right)^{d-3}  + \dots,
~~~
w(r)=\frac{\bar j}{r}\left( \frac{\ell_c }{r } \right)^{d-2} + \dots~,
\label{exp_inf}
\end{eqnarray} 
while the near horizon expansion of the nonextremal solutions  is 
 \begin{eqnarray}
\label{c1}
&&f(r)=f_1(r-r_h)+  O(r-r_h)^2,~~h(r)=h_h+ O(r-r_h),
\\
\nonumber
&&
b(r)=b_1(r-r_h)+O(r-r_h)^2,~~w(r)=w_h+ O(r-r_h),
\end{eqnarray}
with $f_1,b_1,h_h$ positive constants.
 For the solutions 
within the ansatz (\ref{metric-rot}), the 
event horizon's angular velocities are 
all equal, $\Omega_k=\Omega_H=w(r)|_{r=r_h}$.
 The Killing vector  $\chi=\partial/\partial_t+
\sum_k\Omega_k \partial/\partial \varphi_k $ is 
orthogonal to and null on the horizon.

The  Hawking temperature  and the event horizon area of these configurations are
\begin{eqnarray} 
\label{Temp-rot} 
  T_H=\frac{\sqrt{b'(r_h)f'(r_h)}}{4\pi},~~~~A_H= r_h^{d-3} \sqrt{h_h}{ V}_{1,d-2} .
\end{eqnarray} 
The conserved 
charges  of the rotating black holes are obtained by 
using  again the counterterm  method in conjunction 
with the quasilocal formalism.  
Based on the results we have obtained for spacetime dimensions $d=5,7,9$, 
we propose the
following general expression  for mass-energy  and 
angular momentum\footnote{Note that these quantities are 
evaluated in a frame which is non-rotating
at infinity.}:  
\begin{eqnarray}
\label{MJ-rot} 
M= \frac{\ell_c^{d-3} U}{16\pi G 
}\big[\bar f-(d-1)\bar b  \big] V_{1,d-2}+M_1^0~,~~~
 J_{(k)}=J=\frac{\ell_c^{d-2}}{8\pi G_d }\bar j UV_{1,d-2}~~,
\end{eqnarray} 
with $M_1^0$ the Casimir energy as given by (\ref{Casimir}).
The entropy of these solutions as computed from the general relation (\ref{GibbsDuhem}) 
 (with $\mu_i=J,~~{\frak C}_{i}=\Omega_H$) is
\begin{eqnarray}
\label{S-rot} 
S=S_0+S_{c},~~{\rm with}~~S_0=\frac{A_H}{4G},~~
S_{c}=\alpha\frac{ V_{1,d-2}}{8G}(d-3)\sqrt{h_h}r_h^{d-5}(d-1-\frac{h_h}{r_h^2}).
\end{eqnarray} 
It is interesting to note that the entropy (\ref{S-rot}) can also be written in the Wald's form (\ref{S-Noether}). 
However, the derivation of (\ref{S-Noether}) in Ref. \cite{Clunan:2004tb} covers the case of static solutions only; 
it would be interesting to extend it by including the effects of rotation.
Of course, the ultimate test of the formulae (\ref{MJ-rot}), (\ref{S-rot}) 
will be possible when the exact solutions  will be found ($i.e.$ the relation between $b_1,f_1,h_h$ and $\bar f, \bar b,\bar j$) and verify that, with these definitions,
the first law of thermodynamics is satisfied.

The boundary metric upon which the dual field theory resides
corresponds to a  static Einstein universe in $(d - 1)$ dimensions with line element
$h_{ab}dx^adx^b= \ell_c^2 d\Sigma^2_{1,d-2}-dt^2$.
The
stress tensor for the boundary dual theory has also an interesting form. 
Restricting to $d=5$, one finds  (with $x^1=\theta_1,~x^2=\varphi_1,~ x^3=\varphi_2,~x^4=t$ and
$d\Sigma^2_{1,3}=d\theta_1^2+\sin^2 \theta_1\varphi_1^2+\cos^2 \theta_1\varphi_2^2$)
\begin{eqnarray}
\nonumber
8 \pi G  <\tau^{a}_b> =&&
(
\frac{3U-2}{8\ell_c} -\frac{\bar fU}{2\ell_c}
)
\left( \begin{array}{cccc}
1&0&0&0
\\
0&1&0&0
\\
0&0&1&0
\\
0&0&0&-3
\end{array}
\right)
-
\frac{2U}{ \ell_c}(\bar b-\bar f) 
\left( \begin{array}{cccc}
0&0&0&0
\\
0&\sin^2 \theta_1&\cos^2 \theta_1&0
\\
0&\sin^2 \theta_1&\cos^2 \theta_1&0
\\
0&0&0&-1
\end{array}
\right)
\\
\label{st1}
&&+
2 \hat j U \left( \begin{array}{cccc}
0&0&0&0
\\
0&0&0&\sin^2 \theta_1
\\
0&0&0&\cos^2 \theta_1
\\
0&-\frac{1}{\ell^2_c}&-\frac{1}{\ell^2_c}&0
\end{array}
\right).
\end{eqnarray}
As expected, this traceless stress tensor  is finite and covariantly 
conserved.

\section{EGB-AdS black string solutions}
The $k=1$ Schwarzschild-AdS(-GB) black hole solutions in $d$-dimensions
have an event 
horizon of topology $S^{d-2}$,  
which matches the $S^{d-2}$ topology of the spacelike infinity.
Horowitz and Copsey found in  \cite{Copsey:2006br}  a different type of $d=5$ solution of 
Einstein equations with negative $\Lambda$, 
with an event horizon topology $S^{2}\times S^1$. The  configurations there have no 
dependence on the `compact' extra dimension, 
and their conformal boundary is the product of 
time and $S^{2}\times S^1$.
These solutions have been generalized to higher 
dimensions $d\geq5$ in \cite{Mann:2006yi},  
configurations with an event horizon topology 
$H^{d-3}\times S^1$ being considered as well. 

Black objects with event horizon topology 
$S^{d-3}\times S^1$ matching that of the 
spacelike infinity are familiar from the
 $\Lambda=0$ physics and they are usually 
called black strings \cite{bs}. 
The solutions in \cite{Copsey:2006br, Mann:2006yi}
present many similar properties with the  $\Lambda=0$ case,
and are naturally interpreted as the AdS counterparts of these 
configurations
\footnote{Note that these 
configurations 
are very different from the warped AdS solutions as discussed for instance in \cite{Chamblin:1999by}, 
although the latter are also usually called black strings in the literature.
 General remarks about the
properties of these solutions in Lovelock gravity are presented in \cite{Kastor:2006vw}.}. 
Although the AdS black strings are not known in closed form\footnote{See, 
however, the  $d=5$ supersymmetric Einstein-U(1) magnetic black string exact
solutions in \cite{Chamseddine:1999xk}.},
one can analyse their
properties by using a combination of analytical and numerical
methods, which is enough for most purposes.

Different from the $\Lambda=0$ limit, it was found in \cite{ Mann:2006yi} that the AdS black 
string solutions  with an event horizon 
topology $S^{d-3}\times S^1$ have a nontrivial, vortex-like
globally regular limit with zero event 
horizon radius.
As argued in \cite{Copsey:2006br, Mann:2006yi}, these solutions
provide the gravity dual of a field theory  on
a $S^{d-3}\times S^1\times S^1$ (or $H^{d-3}\times S^1\times S^1$)
background.

Generalizations of AdS black strings with gauge fields are studied in 
\cite{Chamseddine:1999xk}, \cite{Brihaye:2007vm}, \cite{Brihaye:2007jua}, \cite{Bernamonti:2007bu}. 
The Ref. \cite{Brihaye:2007vm} discussed  also the properties of a set of rotating solutions.
The issue of Gregory-Laflamme instability \cite{Gregory:1993vy}
for AdS black strings in Einstein  gravity
was addressed in \cite{Brihaye:2007ju}, 
nonuniform solutions ($i.e.$ with dependence on the compact 
`extra'-dimension) being constructed at the perturbative level in \cite{Delsate:2008kw}.

In this Section we present arguments for the existence of asymptotically AdS black strings
in EGB theory and analyse their basic properties.
This provides also an interesting aplication of the formalism in Section 2, since the
 AdS black strings have no obvious background.
For $d=5$, the work here generalizes for $\Lambda<0$  the Kaluza-Klein
EGB black strings  discussed in ref. \cite{Kobayashi:2004hq}.

\subsection{The metric ansatz and reduced action}
We consider the following parametrization of the $d$-dimensional line 
element (with $d \geq 5$)
\begin{eqnarray}
\label{metric} 
ds^2=a(r)dz^2+ \frac{dr^2}{f(r)}+g^2(r)d\Sigma^2_{k,d-3}-b(r)dt^2~,
\end{eqnarray}
 the 'extra'-direction $z$ being compact\footnote{For these uniform
 black string solutions, the period $L$ is an arbitrary positive constant and plays no role in 
 our results.
 However, similar to the $\Lambda=0$, its value is crucial
 when discussing the issue of Gregory-Laflamme instability \cite{Brihaye:2007ju}, 
 \cite{Delsate:2008kw} of these objects.} with period $L$. 
 For $g(r)=r$, this reduces to the metric ansatz in \cite{Mann:2006yi}.

Without fixing a metric gauge, a straightforward computation
leads to the following reduced action of the system
 \begin{eqnarray}
\label{Leff} 
A_{eff}=\int dr dt ~L_{eff},~~~{\rm with~~~~} L_{eff}=L_E+\frac{\alpha}{4}L_{GB},
\end{eqnarray}
where
 \begin{eqnarray}
\label{LE} 
\nonumber
&L_E=(d-3)(d-4)\sqrt{\frac{ab}{f}}g^{d-5}(k+fg'^2)+\frac{1}{2}\sqrt{\frac{f}{ab}}g^{d-3}a'b'
+(d-3)\sqrt{abf}g^{d-4}(\frac{a'}{a}+\frac{b'}{b})g'-2 \Lambda g^{d-3}\sqrt{\frac{ab}{f}},~~{~~~~~~~~~~}
\end{eqnarray}
 \begin{eqnarray}
 \nonumber
L_{GB}&=&(d-3)(d-4)\sqrt{\frac{f}{ab}}g^{d-5}a'b'(k-fg'^2)
-\frac{2}{3}(d-3)(d-4)(d-5) \sqrt{abf}g^{d-6}(\frac{a'}{a}+\frac{b'}{b})g'(fg'^2-3k)
\\
\nonumber
& &+\frac{1}{3}(d-3)(d-4)(d-5)(d-6)\sqrt{\frac{ab}{f}}g^{d-7}(3k^2+6kfg'^2-f^2g'^4).
\end{eqnarray}
 The corresponding equations for the metric functions $a,b,f$ are found by taking the variation
 of $A_{eff}$ with respect to $a,b,f$ and $g$ and fixing after that the metric gauge $g(r)=r$ (this is equivalent to directly
 solving the EGB equations, but technically  simpler).
 The resulting relations are very long and we do not include them here.
Similar to the case of Einstein gravity, the equations for $a$ and $f$ are first order while the equation
for $b$ is second order.

\subsection{The asymptotics}
We consider non-extremal black string solutions presenting the following expansion
near the event horizon (taken at constant $r=r_h$) 
\begin{eqnarray}
\label{eh} 
a(r)=a_h+O(r-r_h),
~
b(r)=b_1(r-r_h)+O(r-r_h)^2,~f(r)=f_1(r-r_h)+O(r-r_h)^2,{~~~~}
\end{eqnarray}
with all coefficients fixed by the parameters $a_h$, $b_1$.
The condition for a regular 
event horizon is $f'(r_h)>0$, $b'(r_h)>0$ and $a_h$ a positive constant. In the $k=-1$ case, this 
implies the existence of a minimal value of the event horizon radius.

 We consider solutions of the   equations of motion whose boundary topology  
is the product of time and $S^{d-3}\times S^1$, $R^{d-3}\times S^1$ or $H^{d-3}\times 
S^1$. For even values of the spacetime dimension $d$, the solutions of the field equations
admit  at large $r$  a power series expansion of the form:
\begin{eqnarray} 
\nonumber
a(r)&=&(\frac{d-4}{d-3})k+\frac{r^2}{\ell_{c}^2}+\sum_{j=1}^{(d-4)/2}a_j(\frac{\ell_{c}}{r})^{2j}
+c_z(\frac{\ell_{c}}{r})^{d-3}+O(1/r^{d-2}),
\\
\label{even-inf}
b(r)&=&(\frac{d-4}{d-3})k +\frac{r^2}{\ell_{c}^2}+\sum_{j=1}^{(d-4)/2}a_j(\frac{\ell_{c}}{r})^{2j}
+c_t(\frac{\ell_{c}}{r})^{d-3}+O(1/r^{d-2}),
\\
\nonumber
f(r)&=&\frac{k(d-1)(d-4)}{(d-2)(d-3)}+\frac{r^2}{\ell_{c}^2}+\sum_{j=1}^{(d-4)/2}f_j(\frac{\ell_{c}}{r})^{2j}
+(c_z+c_t)(\frac{\ell_{c}}{r})^{d-3}+O(1/r^{d-2}),
\end{eqnarray}   
where $a_j,~f_j$ are constants depending on the index
$k$ and the spacetime dimension only. Specifically, we find
\begin{eqnarray}
\label{inf2}  
&&a_1=\frac{ k^2(4U-1)}{27 U},~f_1=\frac{( 59U-11)k^2 }{216 U}~{\rm for~~}d=6,  
\\
&&a_1=\frac{ k^2( 49 U-9 )}{1125U },~f_1=\frac{ (177 U-17)k^2 }{2250U},
\\
\nonumber
&&a_2=-\frac{ 2k(63-U(1381-8118U)) }{253125 U^2 },~f_2=-\frac{8k (9 - U (178 - 1269 U))}{84375 U^2},
~{\rm for~~}d=8,  
\end{eqnarray}  
their expression becoming more complicated for higher $d$, with no general
pattern becoming apparent.
 
Similar to the $\alpha=0$ case, the corresponding expansion for odd values of $d$ contains logarithmic terms
\begin{eqnarray}
\nonumber 
a(r)&=&(\frac{d-4}{d-3})k+\frac{r^2}{\ell_{c}^2}+\sum_{j=1}^{(d-5)/2}\bar a_j(\frac{\ell_{c}}{r})^{2j}
+\zeta\log(\frac {r}{\ell_{c}}) (\frac{\ell_{c}}{r})^{d-3}
+c_z(\frac{\ell_{c}}{r})^{d-3}+O(\frac{\log r}{r^{d-1}}),
\\
\label{odd-inf}
b(r)&=&(\frac{d-4}{d-3})k+\frac{r^2}{\ell_{c}^2}+\sum_{j=1}^{(d-5)/2}\bar a_j(\frac{\ell_{c}}{r})^{2j}
+\zeta\log (\frac {r}{\ell_{c}}) (\frac{\ell_{c}}{r})^{d-3}
+c_t(\frac{\ell_{c}}{r})^{d-3}+O(\frac{\log r}{r^{d-1}}),
\\
\nonumber
f(r)&=&\frac{k(d-1)(d-4)}{(d-2)(d-3)}+\frac{r^2}{\ell_{c}^2}+\sum_{j=1}^{(d-5)/2}\bar f_j(\frac{\ell_{c}}{r})^{2j}
+2\zeta\log (\frac {r}{\ell_{c}}) (\frac{\ell_{c}}{r})^{d-3}
+(c_z+c_t+c_0)(\frac{\ell_{c}}{r})^{d-3}+O(\frac{\log r}{r^{d-1}}),
\end{eqnarray}   
where, restricting to the first two values of $d$
\begin{eqnarray}
\label{inf31}
c_0=\frac{k^2(1-U)}{36 U}~~~{\rm for }~~d=5, ~~
c_0=\frac{ k^3(35 U-17 )}{3200 U}~~~{\rm for }~~d=7,
\end{eqnarray} 
 and
 \begin{eqnarray}
\label{inf3}
&&\zeta=\frac{  k^2}{12}~~~{\rm for }~~d=5, ~~
\zeta= \frac{k(3-57 U) }{1600U}~~~{\rm for }~~d=7,
\\
&&\bar a_1=\frac{ k^2(23U-5)}{320 U},~~~~
\bar f_1=\frac{ k^2(52U-7)}{400 U}~~~{\rm for }~~d=7.
\end{eqnarray}  
For any value of $d$, the large $r$ form of the solutions presents two arbitrary parameters $c_t$ and $c_z$
which  uniquelly fix  the coefficients of all terms decaying faster
than $ {1}/{r^{d-3}}$. 
  
\subsection{The conserved charges and entropy}

Apart from the mass-energy $M$, these solutions possess a second 
 charge associated with the compact $z$ direction, corresponding 
 to the black string's tension ${\mathcal T}$. 
The computation of the boundary stress tensor $T_{ab}$ based on the relations in Section 2 is straightforward and we find the 
following expressions for mass and tension (the relations here extrapolate the results found for
$5\leq d\leq 9$)  
\begin{eqnarray}
\label{MT} 
M&=&M_0+M_c^{(k,d)}~,~~M_0=\frac{\ell^{d-4}_c}{16\pi G 
}\big[c_z-(d-2)c_t\big]U LV_{k,d-3}~,
\\
{\mathcal T}&=&{\mathcal T}_0+{\mathcal T}_c^{(k,d)}~,~~
{\mathcal T}_0=\frac{\ell^{d-4}_c}{16\pi G }\big[(d-2)c_z-c_t\big] 
U V_{k,d-3}~,
\end{eqnarray}  
where
$M_c^{(k,d)}$ and ${\mathcal T}_c^{(k,d)}$ are  Casimir-like terms
which appear for an odd spacetime dimension only,
 \begin{eqnarray}
\label{MT-Cas} 
 M_c^{(k,d)}=-L{\mathcal T}_c^{(k,d)} &=&\frac{\ell^{d-4}_c}{8\pi G 
}V_{k,d-3}  L\bigg(\frac{3-2U}{24}k^2\delta_{d,5}-\frac{10+U(123U-7)}{3200}k\delta_{d,7}
\\
&&
\nonumber
+
\frac{ U (75920 + U (-284038 + 2262727 U))  -1484 }{64012032 U^2}k^2\delta_{d,9}+
\dots\bigg)~.
\end{eqnarray}  
The Hawking temperature and the event horizon area of the black strings  are
  \begin{eqnarray}
T_H=\frac{\sqrt{b'(r_h)f'(r_h)}}{4\pi},~~ A_H=r_h^{d-3}V_{k,d-3}L\sqrt{a_h}.
\end{eqnarray}
To evaluate the black string's action, it is important to use the observation that one can write
\begin{eqnarray}
\label{totder1} 
 &R_t^t+\frac{\alpha}{4}(H_t^t+\frac{1}{2}L_{GB})=
\frac{1}{ r^{d-2}} \sqrt{\frac{f}{ ab }}\frac{d }{dr}
 \left  ( 
  -\frac{1}{2}r^{d-5}b'\sqrt{\frac{af}{ b }}(r^2-
  \frac{1}{2}\alpha(d-3)
  (
  (d-4)(f-k)+f \frac{ra'}{a}
  )
\right ),~~~~{~~} 
\\
\nonumber
 &R_z^z+\frac{\alpha}{4}(H_z^z+\frac{1}{2}L_{GB})=
\frac{1}{ r^{d-2}} \sqrt{\frac{f}{ ab }}\frac{d }{dr}
 \left  ( 
  -\frac{1}{2}r^{d-5}a'\sqrt{\frac{bf}{a }}(r^2-
  \frac{1}{2}\alpha(d-3)
  (
  (d-4)(f-k)+f \frac{rb'}{b}
  )
\right )~.~~~{~~} 
\end{eqnarray} 
Integrating the first relation above taken  together with the field equations 
(\ref{eqs}),  we isolate the bulk action
contribution at infinity and at  $r=r_h$. The divergent 
contributions given by the surface integral term at infinity are 
canceled by $I_{b}^{(E)}+I_{b}^{(GB)}+I_{\text{ct}}$ (with $I_{\text{ct}}=I_{\text{ct}}^0+I_{\text{ct}}^s)$,
which results in a finite expression of $I_{cl}$. 
Together with 
(\ref{GibbsDuhem}), we find the entropy of solutions
\begin{eqnarray}
\label{S-UBS} 
S= S_0+S_c,~~~{\rm with}~~~S_0=\frac{A_H}{4G}~,
~~~~S_c=\frac{k}{4G}\frac{\alpha}{2}(d-3)(d-4)r_h^{d-5}LV_{k,d-3}\sqrt{a_h}~.
\end{eqnarray}
Therefore, similar to the black hole case, the entropy  of a black string is not simply proportional
to the black hole horizon area as it is in the Einstein gravity, but has an extra term
proportional to the GB coupling parameter (note that, as expected, the GB contribution to $S$
is proportional to the extra-term in the entropy of a ($d-1$)-dimensional Schwarzschild-GB-AdS black hole (\ref{SGB})).
Moreover, the entropy of a black string can also be formally written in Wald's form (\ref{S-Noether}), the GB contribution being proportional
again with the Ricci scalar of the event horizon. 

By using the second relation in (\ref{totder1}) we find also
that
$I =-\beta {\mathcal T}L$.
This relation together with (\ref{GibbsDuhem})   leads to an unexpectedly simple 
Smarr-type formula, relating quantities defined at 
infinity to quantities defined at the event horizon:
\begin{eqnarray}
\label{smarrform} 
M+{\mathcal T}L=T_H S~,
\end{eqnarray} 
which is the result found in \cite{Mann:2006yi} for solutions without a GB term.
This relation also provided a useful check of the 
accuracy of the numerical solutions we have obtain. 

 The first law of thermodynamics (\ref{1stlaw}) looks more complicated for black strings as compared to the black hole case,
since the lenght scale $L$ enters there 
as an extensive parameter \cite{Townsend:2001rg}, 
$i.e.$ $\mu={\mathcal T}$, ${\frak C}=L $.
In the absence of closed form solutions, the validity of 
the generic relation (\ref{1stlaw})
can be verified only numerically. 
In practice, for both $d=5$ and $d=6$ solutions, 
we have integrated the first law for a fixed period $L$
and computed a value of mass.
Then the expression of the tension is computed\footnote{This is the standard approach used to compute the 
mass and tension of the $\Lambda=0$ black string solutions presenting a dependence
on the extra-coordinate $z$, in which case it has proven very difficult to read $M$ and ${\mathcal T}$ 
from the asymptotic data at infinity, see $e.g.$ \cite{Kudoh:2004hs}.} from the Smarr-type formula (\ref{smarrform}).
The values of $M$ and ${\mathcal T}$ computed in this way were found to coincide with a reasonable
 accuracy with those derived by using the expression 
(\ref{MT}), up to the overall Casimir terms $M_0$, ${\mathcal T}_0$.

We give here also the expectation value of the stress tensor of the dual theory 
 for the simplest case $d=5$ (with the background metric upon which the dual field theory resides
 $h_{ab} dx^a dx^b=\ell_c^2d\Sigma_{k,2}^2+dz^2-dt^2$, and $x^1=\theta,x^2=\phi,x^3=z,x^4=t$, 
 while $\theta,\phi$ are the coordinates on a surface of constant $r,t$)
\begin{eqnarray}
\nonumber
8 \pi G  <\tau^{a}_b> =&&
k^2
\left( \begin{array}{cccc}
\frac{3-U}{24\ell_c}&0&0&0
\\
0&\frac{3-U}{24\ell_c}&0&0
\\
0&0&\frac{2U-3}{24\ell_c}&0
\\
0&0&0&\frac{2U-3}{24\ell_c}
\end{array}
\right)
+\left( \begin{array}{cccc}
-\frac{(c_t+c_z)U}{2\ell_c}&0&0&0
\\
0&-\frac{(c_t+c_z)U}{2\ell_c}&0&0
\\
0&0& \frac{(3c_z-c_t)U}{2\ell_c}&0
\\
0&0&0&\frac{(3c_t-c_z)U}{2\ell_c}
\end{array}
\right) .
\end{eqnarray}
As expected, this anisotropic perfect fluid stress tensor  is finite and covariantly 
conserved. However, for $k\neq 0$ it is $not$ traceless, with
\beqs
\label{tikd5dual}
8\pi G<\tau^a_a>= 
\frac{\ell_c^3 }{8}
\bigg(
 (1-\frac{4U}{3})\mathsf{R}^2+(5U-4)\mathsf{R}_{ab}\mathsf{R}^{ab} 
 +(1-U)\mathsf{R}_{abcd}\mathsf{R}^{abcd}
\bigg),
\eeqs
(where the geometric quantities are computed for the metric $h_{ab}$).
For $U=1$ ($i.e.$ $\alpha=0$), this trace is 
equal to the conformal anomaly of the boundary CFT \cite{Skenderis:2000in}.
A similar computation performed for the $d=7,9$ cases leads 
again 
to a nonvanishing trace of the boundary stress tensor. 
As discussed at length in  \cite{Brihaye:2007vm},  the 
trace of $<\tau^a_a>$  for the seven-dimensional Einstein gravity 
black string solutions matches precisely the 
conformal anomaly of the dual 
six-dimensional superconformal $(2,0)$ theory 
\cite{Skenderis:2000in,Bastianelli:2000hi}. 

One can see that the coupling constant $\alpha$ enters the expression of the trace anomaly and Casimir energy
for both EGB black holes and black strings. This is what we expect on general grounds\footnote{However, a more realistic string theory
computation would require to consider solutions in a different model
than (\ref{action}), containing  $\alpha^3$ corrections consisting in
Weyl-squared terms.}, since in EGB theory 
the effective radius (\ref{lc}) of the AdS space and the boundary metric depend  on $\alpha$.

We close this part by remarking that by performing the double analytic
continuation   $z \to i u $, $t \to i\chi$ with a simultaneous
exhange of corresponding charges, the
black strings become static  bubble of nothing solutions in EGB theory.
In order to obtain a regular solution, the spatial coordinate $\chi$ is identified with a period
$\beta = 1/T_H$.
%
\begin{figure}[ht]
\hbox to\linewidth{\hss%
	\resizebox{7cm}{5.8cm}{\includegraphics{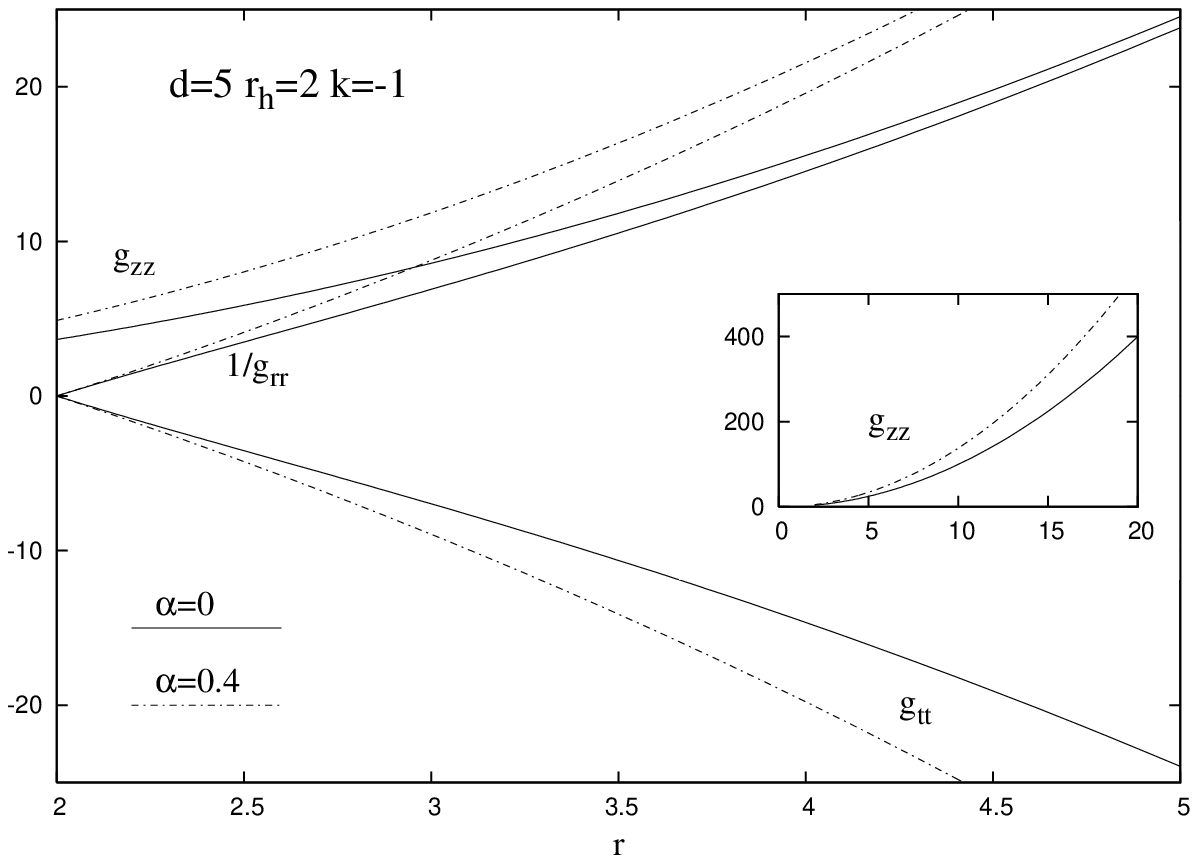}}
\hspace{5mm}%
        \resizebox{7cm}{6.1cm}{\includegraphics{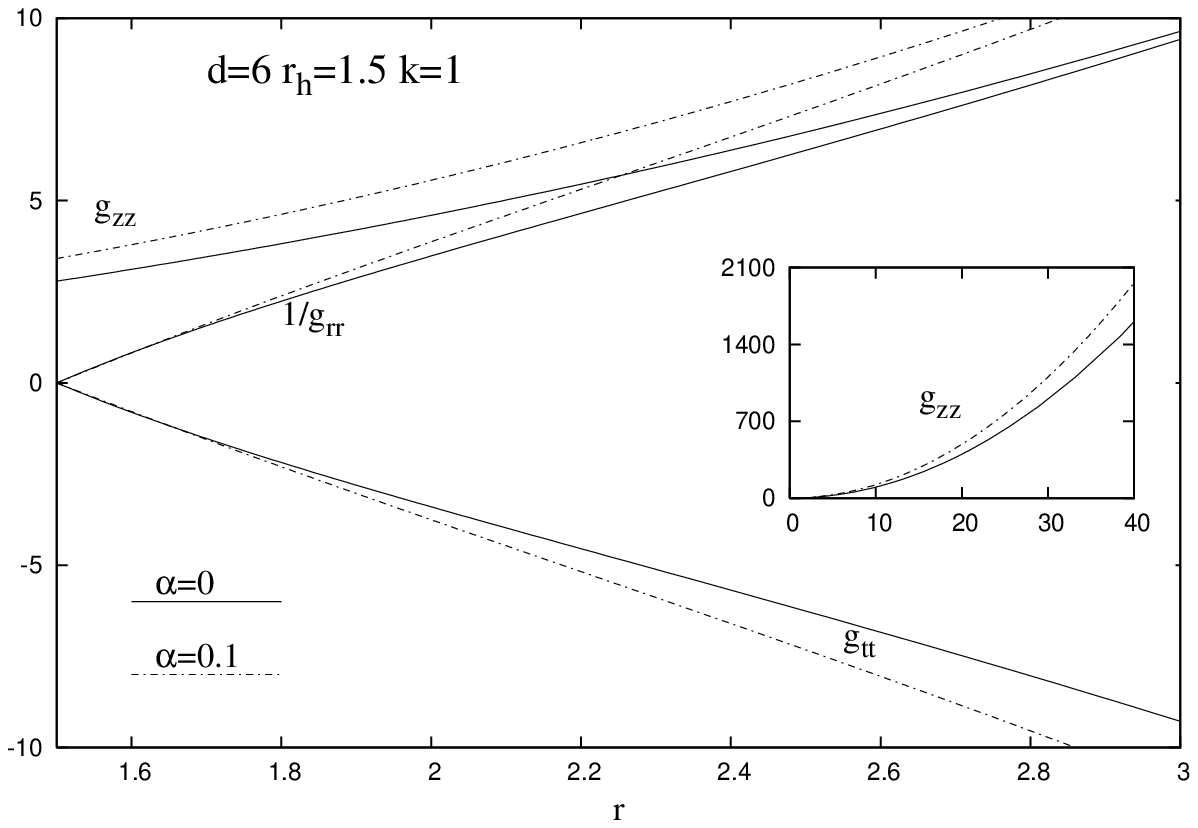}}	
\hss}
	\caption{  The profiles of the metric functions $g_{tt}$, $g_{zz}$ and 
$1/g_{rr}$ are shown for typical $k=\pm 1$ black string solutions in EGB theory. 
For comparison, we included also the profiles of the corresponding solutions in Einstein gravity. 
} 
\label{Fig1}
\end{figure}
\subsection{Numerical results}
In the absence of closed form solutions,
we relied in this case on numerical methods to solve the EGB equations. 
The 
numerical methods here are similar to those  used in literature to study
other $\Lambda<0$ black string solutions  \cite{Brihaye:2007vm}, \cite{Brihaye:2007ju}, \cite{Delsate:2008kw}.
Taking units such that $G=1$,  we used a standard solver  which involves a Newton-Raphson method
  for 
boundary-value ordinary
differential equations, equipped with an adaptive mesh selection procedure
\cite{COLSYS}.
Typical mesh sizes include $10^3-10^4$ points.
The solutions have a typical relative accuracy of $10^{-8}$.

For $k=0$, the EGB equations admit the exact solution $a=r^2$, 
$b=f$ (where $f$ is given by (\ref{f-BH})),
  which was recovered by our numerical 
procedure. This  solution is likely to be unique, corresponding to the 
known planar topological black hole (\ref{metric}).
Therefore  we shall concentrate here on 
the   cases $k=\pm 1$.

In order to construct numerical solutions for a given $d$, the  constants $(\alpha,\Lambda)$ have to be fixed. 
Then the solution is further specified by the event horizon $r_h$.
Given the complexity of the problem, 
the complete classification of the solutions in the space 
of parameters is a considerable task that is 
not aimed in this paper. 
\begin{figure}[ht]
\hbox to\linewidth{\hss%
\resizebox{9.6cm}{7cm}    {\includegraphics{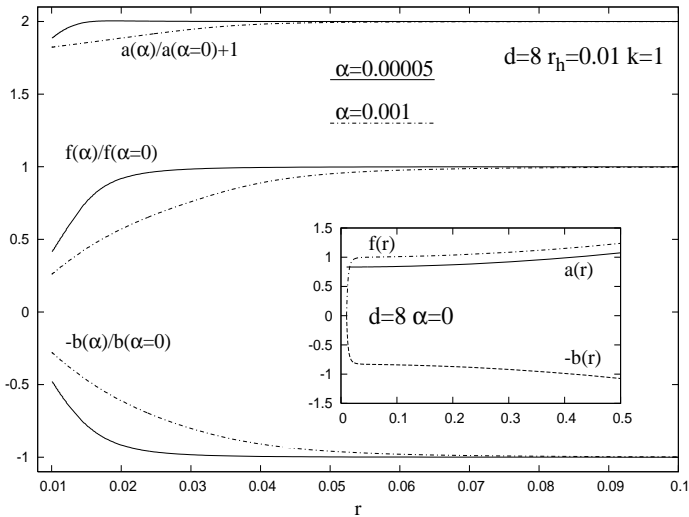}}	 
\hss}  
	\caption{ The effects on the solutions of a nonzero Gauss-Bonnet coupling constant is plotted for two $k=1$
 black strings in $d=8$.
The small $r$ region of the profiles of the metric function in Einstein gravity is also plotted. }
\label{Fig2}
\end{figure}
Instead,  we set $\ell=1$ by using a suitable rescaling, 
and
analyse  in detail a few particular 
classes of solutions, 
which hopefully would reflect all relevant properties 
of the general pattern. 
Also, we shall restrict in this work to the study of the branch of solutions smoothly emerging from the Einstein
gravity configurations. 
Although most of the numerical data presented here corresponds 
to $d=5,6$, in which cases new qualitative features exist, 
we have found also solutions for $d=7,8$.
Therefore we conjecture that they exist for any $d\geq 5$.
 
For all configurations we have studied, the metric functions $a(r)$, $b(r)$ 
and $f(r)$ interpolate monotonically between the corresponding values 
at $r=r_h$ and the asymptotic values at infinity, without presenting 
any local extrema.  
The profiles of the metric functions of  typical EGB  black string solutions
are presented on Figures 1,2
together with the corresponding data in Einstein gravity. 
One can see that a nonzero $\alpha$ leads to a deformation of all metric functions
at all scales
and is  particularly apparent on the function $g_{zz}=a(r)$.

When studying the dependence of the black strings on the GB parameter $\alpha$
and  the event horizon radius $r_h$
we have found that they present an unexpectedly rich structure.
As discussed in \cite{Mann:2006yi}, \cite{Brihaye:2007vm}, the AdS black strings in 
$d$-dimensions follow the general pattern of the  black holes with the same index $k$
in $(d-1)$ dimensions.
This remains valid for EGB configurations. Therefore the cases $d=5,6$ with $k=1$
are rather special, since there are no EGB black holes in four dimensions, while the $d=5$
EGB black holes with spherical horizon have  distinct properties. 

Before discussing the  pattern of solutions, let us briefly recall the structure of the $k=1$ black strings in the
Einstein theory. In \cite{Mann:2006yi} it was shown that a family of black strings exist
presenting a regular horizon at $r=r_h$ for all $r_h >0$. 
On the other hand, the equations also admit a vortex-like regular solution on $r \in [0,\infty)$ with $f(0)=1,~a(0)>0,~b(0)>0$.
\begin{figure}[ht]
\hbox to\linewidth{\hss%
	\resizebox{7cm}{6cm}{\includegraphics{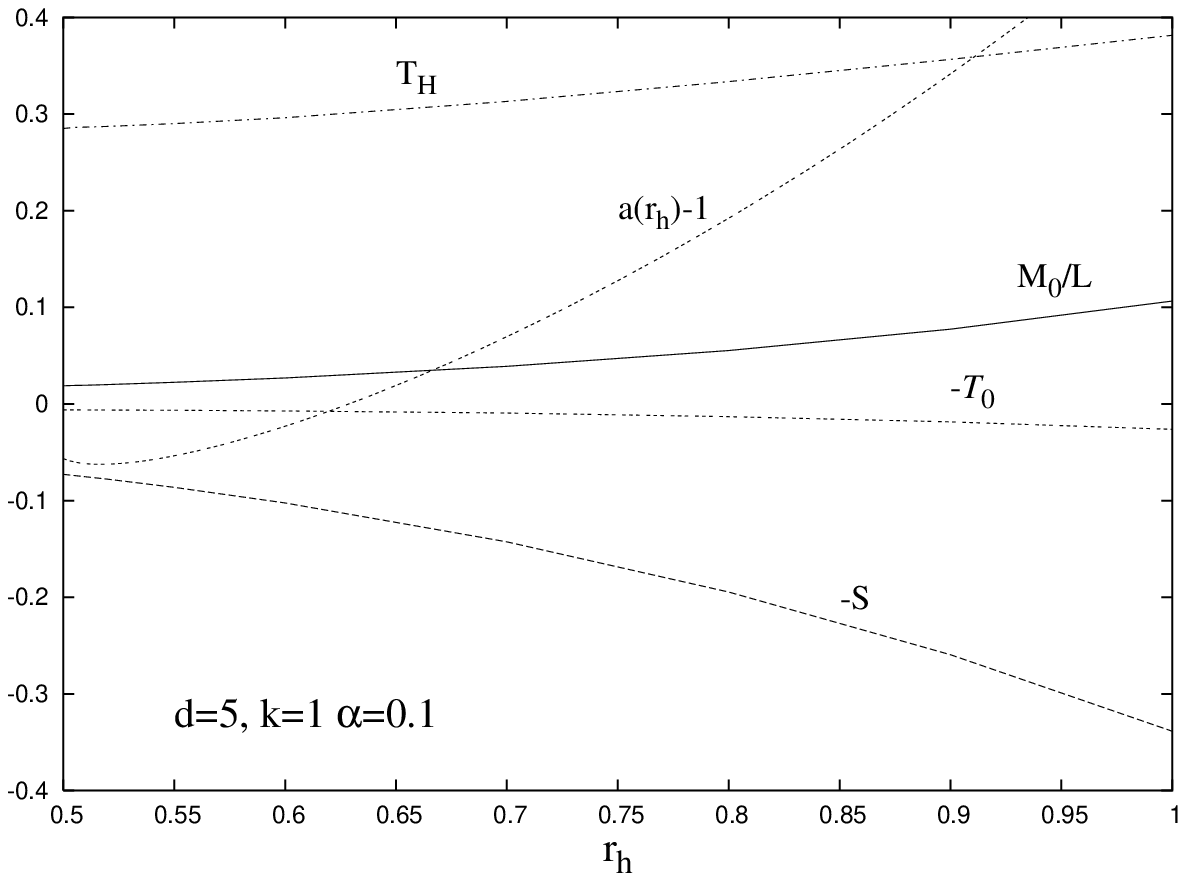}}
\hspace{5mm}%
 \resizebox{7cm}{6cm}{\includegraphics{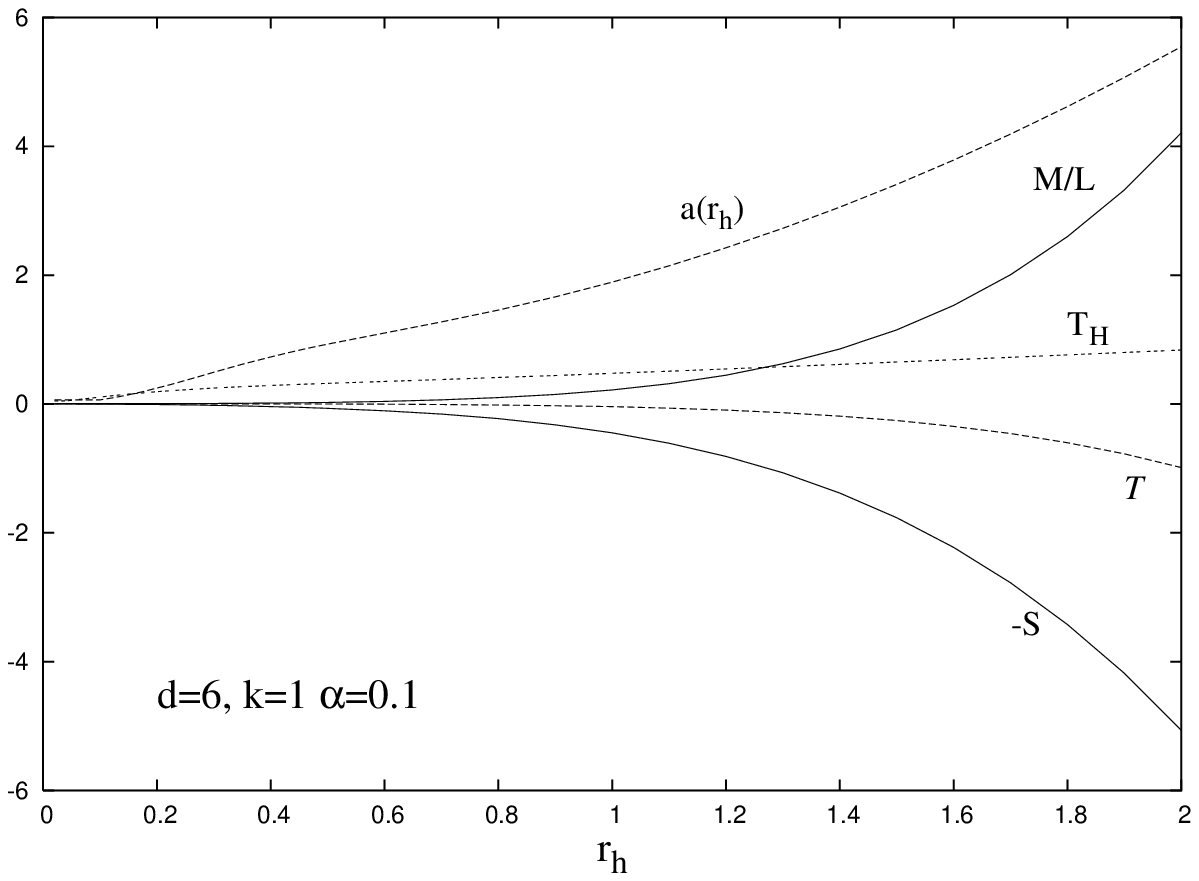}}	
\hss}
	\caption{The mass-parameter $M$, the tension ${\cal T}$ (without the Casimir terms), the value of the
metric function $a(r)$ at the event horizon as well as the Hawking temperature $T_H$ and the entropy
$S$ of $k=1$ black string solutions with $\alpha=0.1$ are represented as functions of the event horizon radius in $d =5,6$ dimensions.}
\label{Fig21}
\end{figure}
 In the limit $r_h \to 0$, the black strings approach this regular solution. 
 In particular, the derivatives
 $f'(r_h)$ and $b'(r_h)$ both diverge for $r_h \to 0$. 
 However, the regular solution has a finite and nozero mass and tension, for any value of $d$.

This picture is very different in the EGB theory.  
First, our numerical results show that
the $d=5,6$ $k=1$ globally regular solutions with $r_h = 0$, which exist for $\alpha = 0$, do not appear
 in the EGB theory.
 For $d=5$ this can be understood by noticing the existence of a 
 minimal allowed value of $r_h$, which results from the expression of the parameter
 $f_1$ in the event horizon expansion (\ref{eh})
 \begin{eqnarray}
\label{d5f1} 
f_1=\frac{r_h^2(\ell^2+2\alpha)}{\ell^2 \alpha}
-\frac{1}{\ell^2 \alpha}\sqrt{(\ell^2-2\alpha)\left(r_h^2(\ell^2-2\alpha)-2\ell^2\alpha \right)}>0,
\end{eqnarray}
(the general $d-$expression is much more complicated),
which implies $r_h>\ell/\sqrt{\ell^2/(2\alpha)-1}$.
Thus there exist a minimal
value of $r_h$ for which the Einstein black strings can be generalized into EGB ones\footnote{As discussed in 
\cite{Kobayashi:2004hq}, the $d=5$ black strings in Kaluza-Klein  theory  ($\Lambda=0$) present also a minimal
horizon size $ r_{h(min)}=\sqrt{2\alpha}$. This can  be seen by taking the 
$\ell\to \infty$ limit in the AdS relation (\ref{d5f1}).}.
This is a nonperturbative effect, which cannot be seen considering the GB term 
as a small deformation  of the Einstein gravity, in which case one finds 
$f_1=1/r_h+4r_h/\ell^2+\alpha/(2r_h^3)+O(\alpha)^2.$

The pattern of $k=1$ solutions is different for $d=6$. Fixing $r_h > 0$, we could construct for all $r_h>0$
a family of solution for $\alpha \in [0, \alpha_{max}]$. In the limit $\alpha \to 0$,
the Einstein theory solution is approached smoothly (different from the $d=5$ case discussed above).
The limit $r_h\to 0$ is different, however, from the $\alpha=0$ case, since no globally regular solution is
found in this case. 
Instead, $a(0)\to 0$ while   $f'(r_h)$ and $b'(r_h)$ take finite (and very small) values in this limit. 
\begin{figure}[ht]
\hbox to\linewidth{\hss%
	\resizebox{7cm}{6cm}{\includegraphics{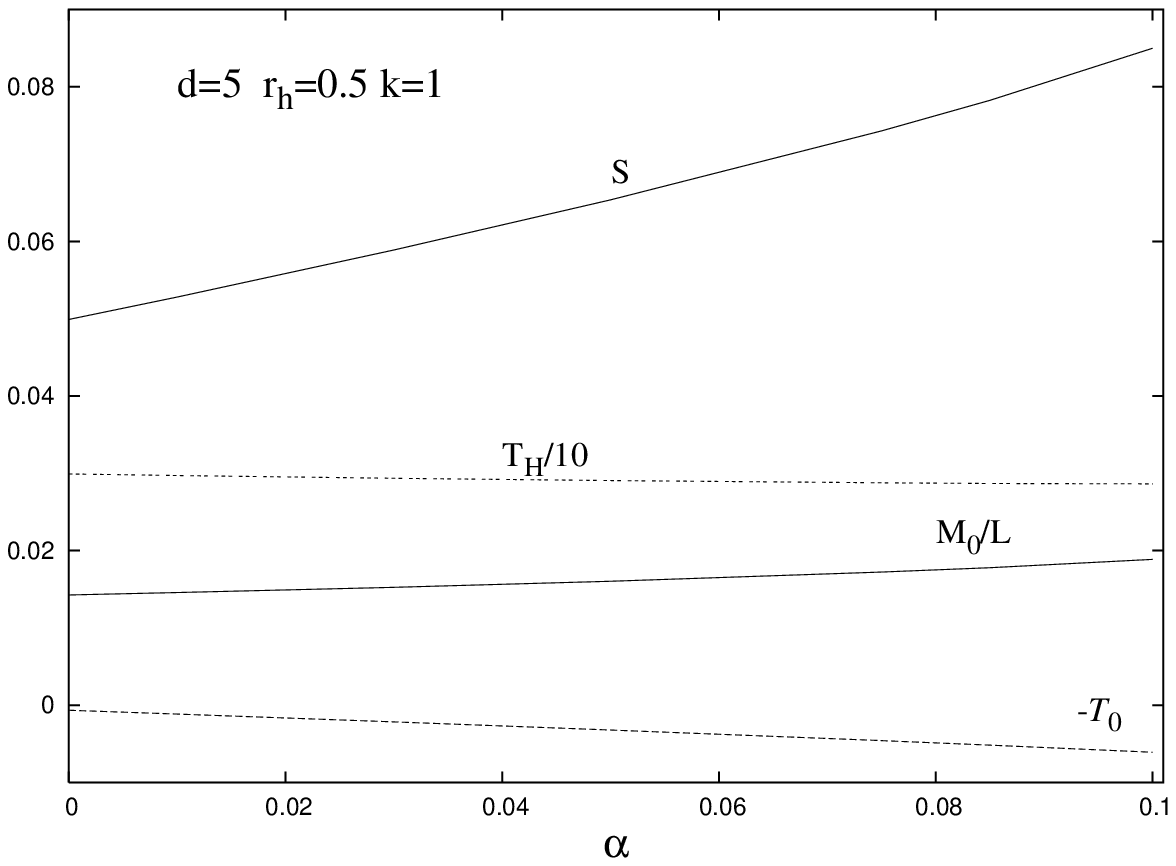}}
\hspace{5mm}%
        \resizebox{7cm}{6cm}{\includegraphics{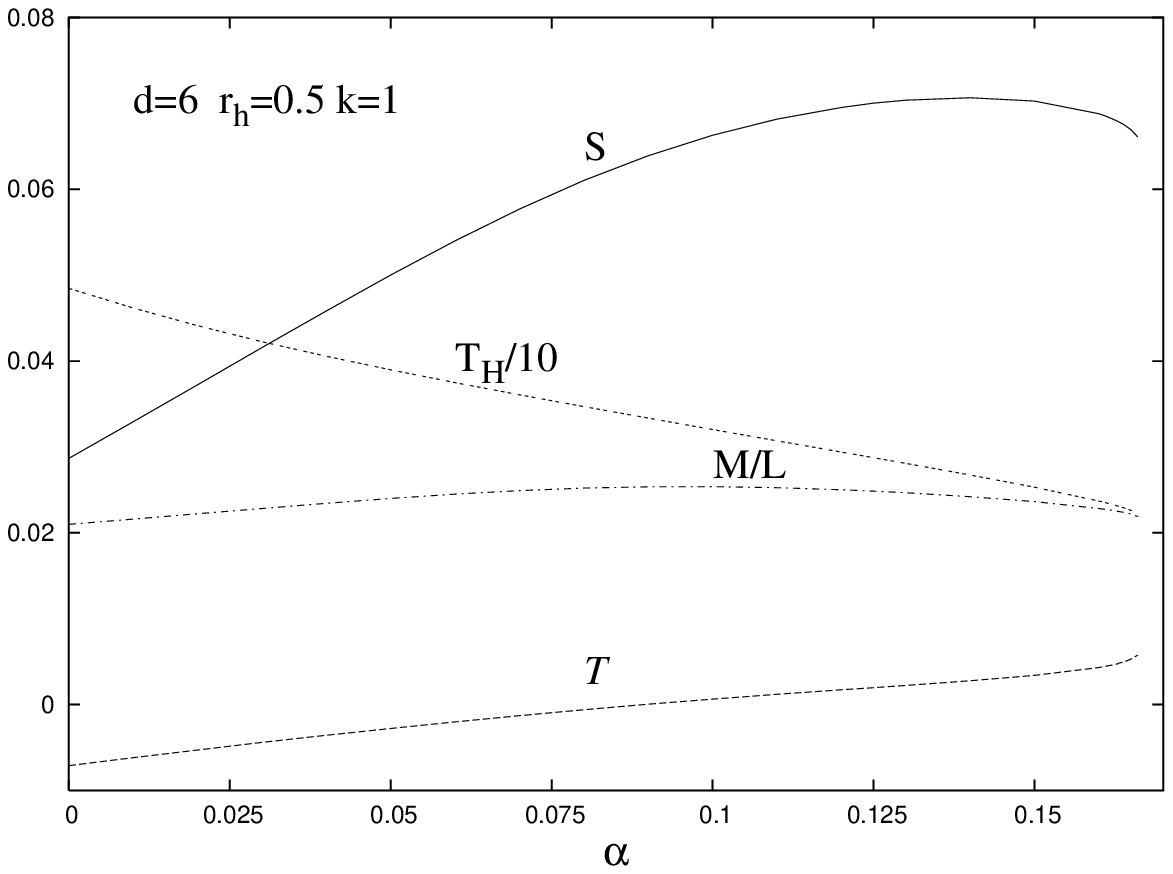}}	
\hss}
	\caption{The Hawking temperature $T_H$, the  mass-energy $M$,  the 
tension ${\cal T}$ and the entropy $S$  are plotted
as a function of  $\alpha$
for $d=5,6$ black string with $k=1$.  
}
\label{Fig22}
\end{figure}
As a result, the Ricci scalar diverges and a naked singular configuration is approached\footnote{
 For both $d=5$ and $d=6$, we have found rather difficult to approach the limiting $r_h$ configurations
 by employing the ansatz (\ref{metric}). A different metric parametrization appears to be necessary.}.
These different observations clearly suggest that the two  limits $\alpha \to 0$ and $r_h \to 0$ do not intercomute
in the pattern of solutions available for $d=6$.

Of course, the results above do not prove the absence of $k=1$ globally regular configurations in EGB theory with 
negative $\Lambda$ (although unlikely, they may be disconnected from the studied black string branch). 
For example, for any $d$ it is possible to write a consistent expansion of the solution near the origin as a power series 
in $r$. However, for both $d=5$ and $d=6$ we have failed to find such solutions, when considering
the corresponding boundary value problem.

The situation appears to be different for $d>6$ (and $k=1$), in which case the $\alpha=0$ pattern found in  \cite{Mann:2006yi}
is still valid.
There the black string event horizon radius is an arbitrary parameter. 
Our numerics suggest that, for a given  $\alpha\leq \alpha_{max}$,
a globally regular solution is approached as 
$r_h\to 0$, while the parameters $f'(r_h)$, $b'(r_h)$ 
diverge in this limit (therefore also the Hawking temperature), while $a(r_h)$ remains finite and nonzero.
These features are presented in Figure 2, where we plot the data for two $d=8$ black string with a small event horizon radius $r_h=0.01$
together with the corresponding solution in Einstein gravity\footnote{One can see that, as implied by (\ref{even-inf}), given the small values of $\alpha$ for the two profiles in Figure 2, the ratio 
between the metric functions in EGB theory and those in 
Einstein  gravity is very close to one for large enough values of the radial coordinate.} 
(the profiles of the metric functions for large
$r_h$ look similar to those in Figure 1 (with $k=1$)).
\begin{figure}[ht]
\hbox to\linewidth{\hss%
	\resizebox{7cm}{6cm}{\includegraphics{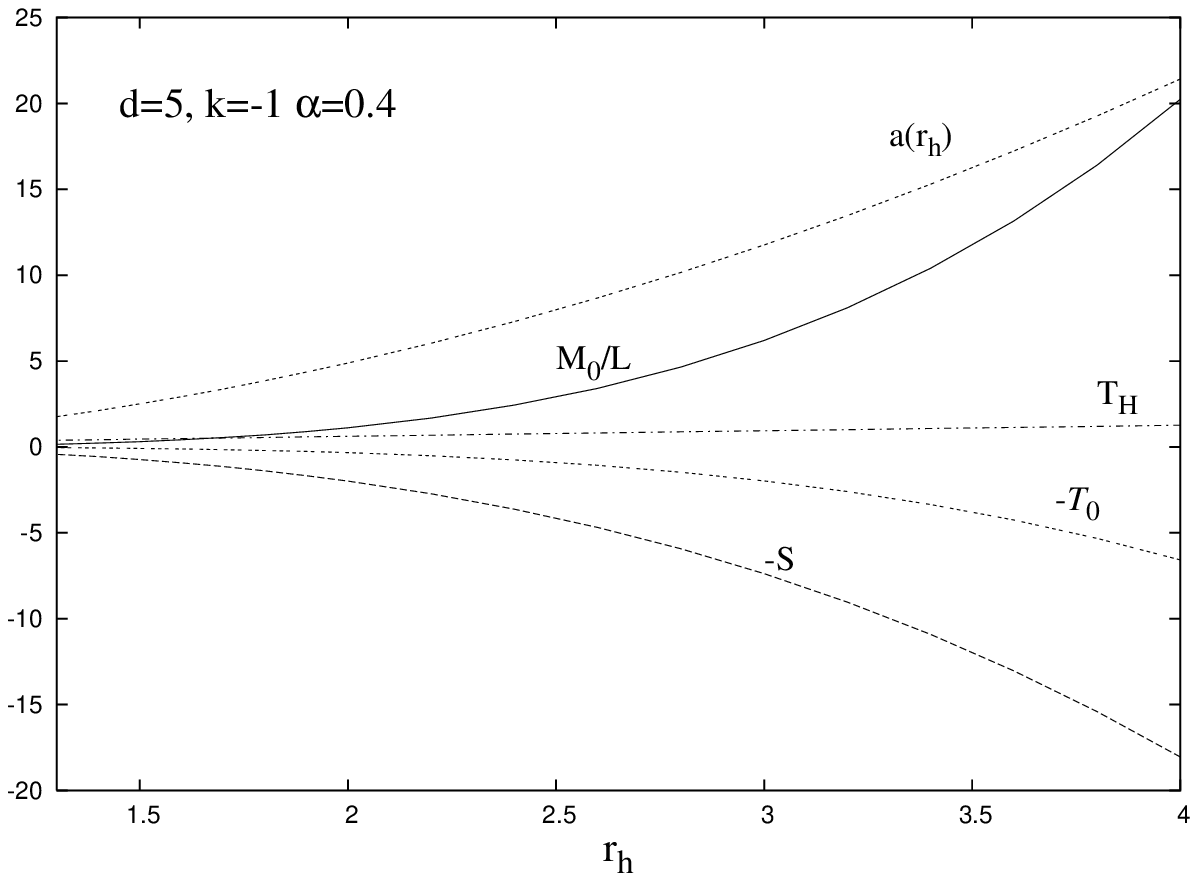}}
\hspace{5mm}%
 \resizebox{7cm}{6cm}{\includegraphics{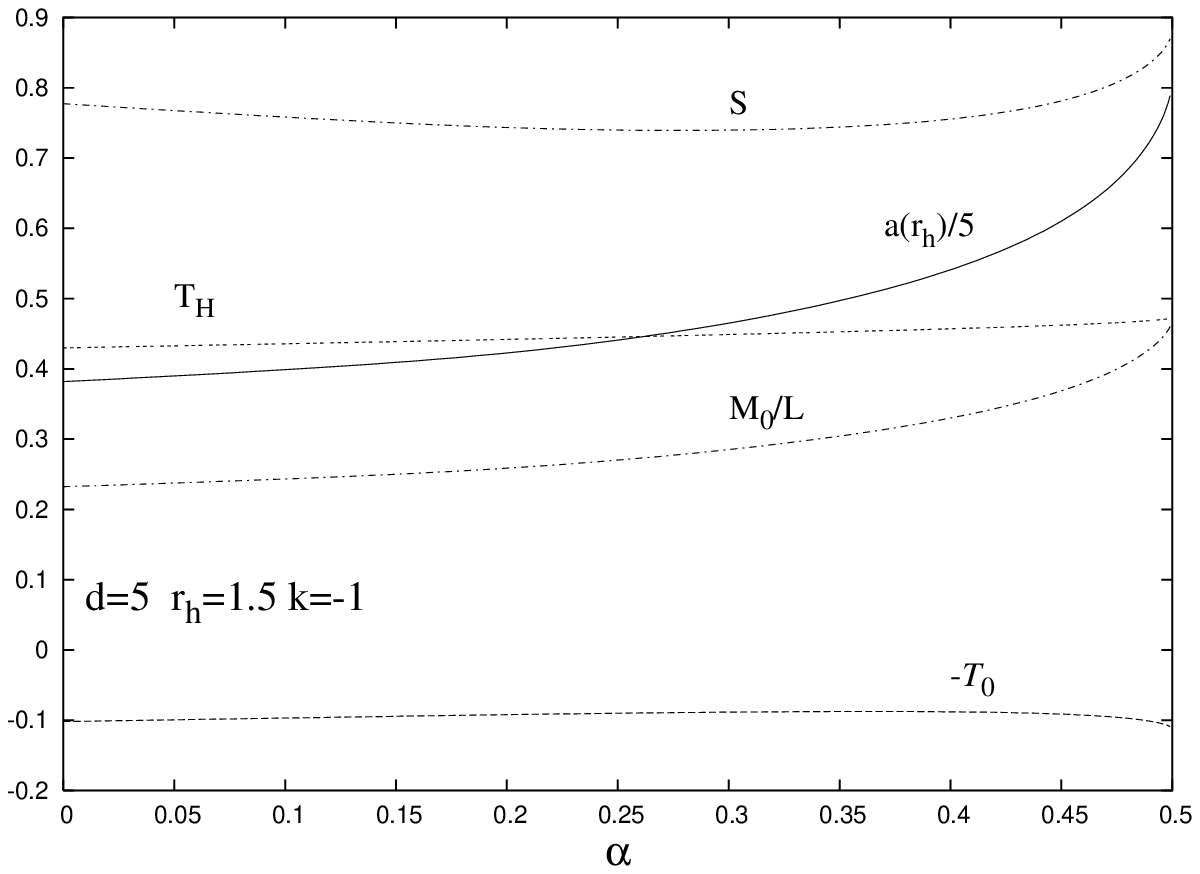}}	
\hss}
	\caption{The dependence of the solutions' parameters on the event horizon radius $r_h$ and Gauss-Bonnet parameter $\alpha$ 
	is presented for $k=-1$ topological black strings in five dimensions. }
\label{Fig25}
\end{figure}
One can see that both $f(r)$ and $b(r)$ present a sharp transition in a small region near the horizon, where
$a(r)$ remain almost constant and nonzero.
However, the severe numerical difficulties encountered for small $r_h$ solutions with $\alpha\neq 0$ prevented us from 
analyzing in details the black string-globally regular vortex
transition. 

We have also increased the event horizon radius $r_h$ for several values of 
$\alpha$ and found
no evidence of a maximal value of $r_h$  where the solutions could eventually
terminate. 
In Figure 3 we plot a number of relevant 
physical quantities as a function of the event horizon radius for the special $d=5$ and $d=6$ cases (the corresponding plots for 
$d=6,7$ present the same qualitative features as the Einstein gravity data exhibited in \cite{Mann:2006yi} and we shall not present
them here).
(Note that all values of $M$, ${\cal T}$ and $S$ plotted in this Section are divided by a factor 
$V_{k,d-2}$.
Also, we have subtracted the Casimir terms from the $d=5$ values of the mass and tension, as computed according to (\ref{MT})).

 We have also studied the dependence of the $k=1$ solutions of the GB parameter $\alpha$, for fixed values of the 
 event horizon radius (see Figure 4).
The solutions exist up to a maximal value of $ \alpha$. 
The configuration  corresponding to $\alpha_{max}$ does not present any special properties.
For those values of $r_h$ where solutions were found,  we stronly suspect the existence of  a second
branch of solutions, also terminating at $\alpha = \alpha_{max}$, 
but it was not attempted to construct it
in a systematic way.

Considering now  topological back string solutions, our results show that
their qualitative features  
are similar to those in the Einstein gravity.
They also exist for all values of $r_h>r_{h(min)}$, 
with $r_{h(min)}$ decreasing with $\alpha$.
 The mass, temperature and entropy of the $k=-1$ configurations increase 
monotonically with $r_h$, while the tension decreases (see Figure 5).
As a result, the thermodynamic of these solutions is similar to the $\alpha=0$ case and they 
present a positive specific heat $C=T_H(\partial S/\partial T_H)>0$, although, from (\ref{S-UBS}) (and similar 
to the black hole case), the entropy may take also negative values.

However, when considering the thermodynamics of black strings,
the situation is much more complicated for $k=1$ solutions.
Restricting to the case of local stability, the behaviour of configurations
 with $\alpha=0$ follows the pattern of the corresponding black hole solutions in AdS background
 \cite{Mann:2006yi}. 
That is
small  black strings have negative specific heat (they are unstable) but large size black strings have positive
specific heat (and they are stable).
\begin{figure}[ht]
\hbox to\linewidth{\hss%
	\resizebox{7cm}{6cm}{\includegraphics{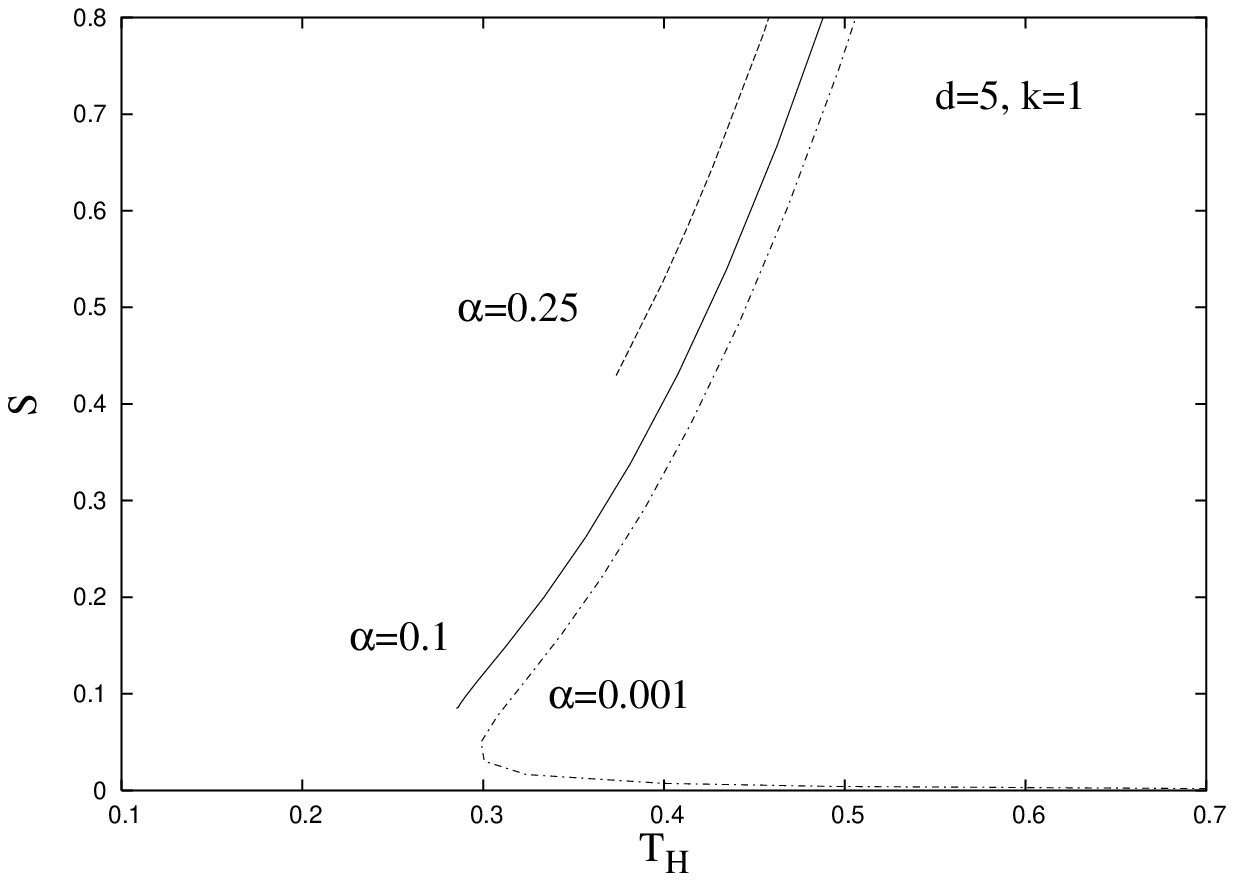}}
\hspace{5mm}%
        \resizebox{7cm}{6cm}{\includegraphics{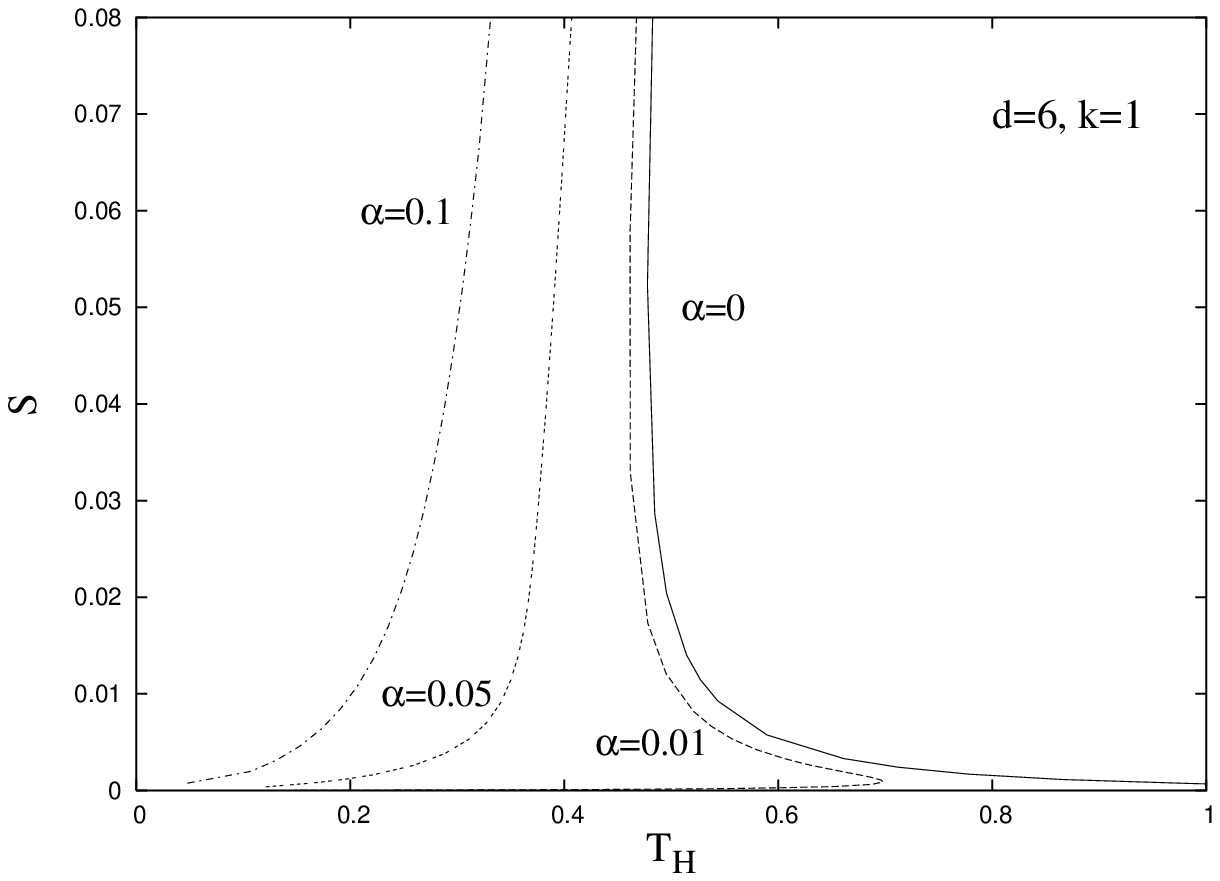}}	
\hss}
	\caption{The   entropy  is plotted
as a function of  Hawking temperature
for $d=5,6$ black string solutions with $k=1$ and several different values of the parameter $\alpha$. }
\label{Fig24}
\end{figure}
However, the picture for $d=6$
is somehow similar to the one of the electrically charged black strings in the canonical ensemble
discussed  in  \cite{Brihaye:2007vm}.
Our results suggest also that  for large enough values of the
parameter $\alpha$  only one branch of stable solutions exist.
As the coupling constant decreases bellow a critical value, additional branches appear, 
of intermediate and small sizes, of which the former has negative
specific heat while the small black strings branch has positive specific heat (see Figure 6).
The picture for $d=5$ is complicated by the existence of a minimal value of the event horizon radius for a given 
value of $\alpha$.
For small enough values of $\alpha$, one finds two branches of solutions, with a negative specific heat for
the small black strings branch.
As seen in Figure 6, if the GB parameter is large enough, all black strings solutions are locally stable.
As expected, for $d>6$, the thermodynamics of the EGB black strings exhibits the same qualitative features as in the
Einstein gravity case.

\section{Further remarks}
The main purpose of this work was to present the boundary counterterm  
that removes the divergences of the action and conserved quantities of the solutions in 
EGB theory with negative cosmological constant for a spacetime dimension $d\leq 9$.
The basic pieces are those used in Einstein gravity plus Lovelock gravity densities. 
Their coefficients, however, present an explicit dependence of the dimensionless factor $\alpha\Lambda$.

One expects that once these coefficients are fixed, one may use the same counterterms
to regulate the action for any choice of coordinates on any asymptotically AdS solution in EGB theory.
This approach is useful particularly for the cases where appropriate backgrounds are ambiguous or unknown.
For example, we have found that the proposed boundary counterterms
provides a finite action and mass for the  NUT-charged solutions in  EGB gravity 
found in \cite{Dehghani:2005zm}.
These configurations are particularly interesting, since as discussed $e.g.$ in \cite{Astefanesei:2004ji},
the usual relationship between area and entropy
is already violated in Einstein theory for a nonzero NUT charge.
The boundary counterterm formalism provides a possibility to evaluate the entropy of such solutions
in theories with higher derivatives.

The counterterm proposed in Section 2 has been derived by considering a range of 
asymptotically AdS solutions in EGB gravity, which does not guarantees its universality. 
However, on general grounds, one expects the boundary counterterm action in EGB theory to be universal, being composed 
of a unique linear combination of curvature invariants that cancel the divergences in the total action
in the limit when the boundary contains the full spacetime. 
For asymptotically AdS solutions in the Einstein gravity, there exist an algorithmic procedure
for constructing $I_{ct}$ in a rigurous way, and so its determination is unique for $\alpha=0$ \cite{Kraus:1999di}.
This procedure involves solving the Einstein equations (written in Gauss-Codacci form) in terms
of the extrinsec curvature functional of the boundary and its derivatives to isolate the divergent parts.
All divergent contributions can be expressed in terms of intrinsic boundary data and do not depend
on normal derivatives.
In principle, this approach can be extended to asymptotically AdS solutions 
  in EGB theory, the only obstacle beng the tremendous complexity of 
the required computation.

In the second part of our paper we have applied the general formalism
to two different kind of
asymptotically AdS black objects in EGB theory.
The results we have found in the static black hole case are similar to those exhibited
in the literature by employing a different approach.
In Section 4 we have presented a set of new solutions,
generalizing the Einstein gravity black strings with $\Lambda<0$
to  EGB theory. 
As argued there, the presence of a GB term in the lagrangean leads to some interesting new features in five and six dimensions, 
in particular
the absence of vortex-like solutions without an event horizon.
The phase structure there gets also modified when including $\alpha-$corrections.
For example, in addition to the large thermodynamically stable black strings,
there are also small $d=6$ stable solutions along with  intermediate unstable ones.
The issue of Gregory-Laflamme instability of these black string is an interesting question.
Based on the Gubser-Mitra conjecture \cite{Gubser:2000ec} that correlates
the dynamical and thermodynamical stability,
we expect  a more complex picture in this case than for the Einstein gravity solutions \cite{Brihaye:2007ju}.

One can address also the situation when  matter fields are added to the bulk action
 (\ref{action}).
One can verify that the counterterms 
proposed in Section 2 regularize \cite{miki} the action and mass of the  Reissner-Nordstrom
generalizations of the black holes (\ref{SGB}). 
However, the situation is different for $d\geq 5$ nonabelian theories \cite{Brihaye:2007jua}, \cite{Radu:2005mj}
or for $d=5$ black strings with a
magnetic U(1) field \cite{Bernamonti:2007bu}, 
in which case new, non-geometric counterterms
should be added to the action already for $\alpha=0$.
An interesting open problem would be to find the corresponding counterterm expression
for $d\geq 7$ asymptotically AdS solutions with higher order terms in the
Lovelock gravity.

\section*{Acknowledgements}
We would like to thank  R. B. Mann for helpful remarks on a draft of this paper.
Y. B. thanks  the
Belgian FNRS for financial support.
 
 \begin{small}

 \end{small}


\begin{thebibliography}{99}
\bibitem{Maldacena:1997re}
J.~M.~Maldacena,
Adv.\ Theor.\ Math.\ Phys.\  {\bf 2} (1998) 231
[Int.\ J.\ Theor.\ Phys.\  {\bf 38} (1999) 1113]
[arXiv:hep-th/9711200].
\bibitem{1} 
  D.~J.~Gross and E.~Witten,
  Nucl.\ Phys.\  B {\bf 277} (1986) 1;
 \\
  R.~R.~Metsaev and A.~A.~Tseytlin,
  Phys.\ Lett.\  B {\bf 191} (1987) 354;
 \\
   C. G. Callan, R. C. Myers, and M. J. Perry, 
   Nucl. Phys. {\bf B311} (1988) 673.  
\bibitem{Myers:1987yn}
  R.~C.~Myers,
  Phys.\ Rev.\  D {\bf 36} (1987) 392.
\bibitem{Grisaru:1986vi}
  M.~T.~Grisaru and D.~Zanon,
  Phys.\ Lett.\  B {\bf 177} (1986) 347;
  \\
  M.~D.~Freeman, C.~N.~Pope, M.~F.~Sohnius and K.~S.~Stelle,
  Phys.\ Lett.\  B {\bf 178} (1986) 199;
  \\
  A.~A.~Tseytlin,
  Phys.\ Lett.\  B {\bf 176} (1986) 92.
   
\bibitem{Fayyazuddin:1998fb}
  A.~Fayyazuddin and M.~Spalinski,
  Nucl.\ Phys.\  B {\bf 535} (1998) 219
  [arXiv:hep-th/9805096].

\bibitem{Aharony:1998xz}
  O.~Aharony, A.~Fayyazuddin and J.~M.~Maldacena,
  JHEP {\bf 9807} (1998) 013
  [arXiv:hep-th/9806159].
  
\bibitem{Nojiri:2000gv}
  S.~Nojiri and S.~D.~Odintsov,
  JHEP {\bf 0007} (2000) 049
  [arXiv:hep-th/0006232].
\bibitem{Balasubramanian:1999re} 
V.~Balasubramanian and P.~Kraus, 
Commun.\ Math.\ Phys.\ \textbf{208}
(1999) 413 [arXiv:hep-th/9902121]. 

\bibitem{Brown:1993br}
J.~D.~Brown and J.~W.~York,
Phys.\ Rev.\ D {\bf 47}  1407 (1993).
\bibitem{Cai:1999xg}
  R.~G.~Cai and N.~Ohta,
  Phys.\ Rev.\  D {\bf 62} (2000) 024006
  [arXiv:hep-th/9912013].
   
\bibitem{kofinas}
  G.~Kofinas and R.~Olea,
  Phys.\ Rev.\  D {\bf 74} (2006) 084035
  [arXiv:hep-th/0606253].
\bibitem{Miskovic:2007mg}
  O.~Miskovic and R.~Olea,
  JHEP {\bf 0710} (2007) 028
  [arXiv:0706.4460 [hep-th]].
\bibitem{Kofinas:2007ns}
  G.~Kofinas and R.~Olea,
  JHEP {\bf 0711} (2007) 069
  [arXiv:0708.0782 [hep-th]].
  
\bibitem{Olea:2006vd}
  R.~Olea,
  JHEP {\bf 0704} (2007) 073
  [arXiv:hep-th/0610230].
    
\bibitem{Cvetic:2001bk}
  M.~Cvetic, S.~Nojiri and S.~D.~Odintsov,
  Nucl.\ Phys.\  B {\bf 628} (2002) 295
  [arXiv:hep-th/0112045].
 \bibitem{Nojiri:2001ae}
  S.~Nojiri, S.~D.~Odintsov and S.~Ogushi,
  Phys.\ Rev.\  D {\bf 65} (2002) 023521
  [arXiv:hep-th/0108172].
\bibitem{Brihaye:2008kh}
  Y.~Brihaye and E.~Radu,
  Phys.\ Lett.\  B {\bf 661} (2008) 167
  [arXiv:0801.1021 [hep-th]].
\bibitem{Dehghani:2006dh}
  M.~H.~Dehghani and R.~B.~Mann,
  Phys.\ Rev.\  D {\bf 73} (2006) 104003
  [arXiv:hep-th/0602243];
\\
  M.~H.~Dehghani, N.~Bostani and A.~Sheikhi,
  Phys.\ Rev.\  D {\bf 73} (2006) 104013
  [arXiv:hep-th/0603058].
\bibitem{Copsey:2006br}
  K.~Copsey and G.~T.~Horowitz,
  JHEP {\bf 0606} (2006) 021
  [arXiv:hep-th/0602003].
\bibitem{Mann:2006yi}
  R.~B.~Mann, E.~Radu and C.~Stelea,
  JHEP {\bf 0609} (2006) 073
  [arXiv:hep-th/0604205].
\bibitem{Lovelock:1971yv}
  D.~Lovelock,
  J.\ Math.\ Phys.\  {\bf 12} (1971) 498.
\bibitem{Mardones:1990qc}
  A.~Mardones and J.~Zanelli,
  Class.\ Quant.\ Grav.\  {\bf 8} (1991) 1545.
\bibitem{GibbonsHawking1}
G.~W.~Gibbons and S.~W.~Hawking,
  Phys.\ Rev.\  D {\bf 15} (1977) 2752.  
\bibitem{Okuyama:2005fg}
  N.~Okuyama and J.~i.~Koga,
  Phys.\ Rev.\  D {\bf 71} (2005) 084009
  [arXiv:hep-th/0501044].
\bibitem{Emparan:1999pm}
  R.~Emparan, C.~V.~Johnson and R.~C.~Myers,
  Phys.\ Rev.\  D {\bf 60} (1999) 104001
  [arXiv:hep-th/9903238].
\bibitem{Mann:1999pc}
  R.~B.~Mann,
  Phys.\ Rev.\  D {\bf 60} (1999) 104047
  [arXiv:hep-th/9903229].
\bibitem{Das:2000cu}
  S.~Das and R.~B.~Mann,
  JHEP {\bf 0008} (2000) 033
  [arXiv:hep-th/0008028].
\bibitem{Davis:2002gn}
  S.~C.~Davis,
  Phys.\ Rev.\  D {\bf 67} (2003) 024030
  [arXiv:hep-th/0208205];
  \\
  E. Gravanis and S. Willison,  Phys.\ Lett.\ B {\bf 562} (2003) 118
  [arXiv:hep-th/0209076].    
\bibitem{MuellerHoissen:1985mm}
  F.~Mueller-Hoissen,
  Phys.\ Lett.\  B {\bf 163} (1985) 106.
\bibitem{Papadimitriou:2004ap}
  I.~Papadimitriou and K.~Skenderis,
  arXiv:hep-th/0404176.
  \bibitem{Myers:1999ps}
  R.~C.~Myers,
  Phys.\ Rev.\ D {\bf 60}, 046002 (1999)
  [arXiv:hep-th/9903203].
\bibitem{Astefanesei:2005ad}
  D.~Astefanesei and E.~Radu,
  Phys.\ Rev.\  D {\bf 73} (2006) 044014
  [arXiv:hep-th/0509144].
\bibitem{quasi}
  J.~D.~Brown, E.~A.~Martinez and J.~W.~York, Phys. Rev. Lett. {\bf 66} 2281 (1991);
  \\
J.~D.~Brown and J.W. York, Phys. Rev. {\bf D47}  (1993) 1407; 
  Phys. Rev. {\bf D47} (1993) 1420  [arXiv:gr-qc/9405024];
    \\
I.~S.~Booth and R.~B.~Mann,
  Phys.\ Rev.\ Lett.\  {\bf 81}  (1998) 5052
  [arXiv:gr-qc/9806015];
    \\
I.~S.~Booth and R.~B.~Mann,
  Nucl.\ Phys.\  B {\bf 539}  (1999) 267
  [arXiv:gr-qc/9806056].
\bibitem{Hawking:ig} S.~W.~Hawking in \textit{General Relativity. An
Einstein Centenary Survey}, edited by S.~W.~Hawking and W.~Israel,
(Cambridge, Cambridge University Press, 1979).   
\bibitem{Mann:2003 Found} 
R.~B.~Mann,
Found.\ Phys.\ \textbf{33} (2003) 65 [arXiv:gr-qc/0211047].
\bibitem{Deser}
D.~G.~Boulware and S.~Deser,
Phys.\ Rev.\ Lett.\  {\bf 55} (1985) 2656.
\bibitem{Cai:2001dz}
  R.~G.~Cai,
  Phys.\ Rev.\  D {\bf 65} (2002) 084014
  [arXiv:hep-th/0109133].
\bibitem{Nojiri:2002qn}
  S.~Nojiri and S.~D.~Odintsov,
  Phys.\ Rev.\  D {\bf 66} (2002) 044012
  [arXiv:hep-th/0204112].
 \bibitem{Cho:2002hq}
  Y.~M.~Cho and I.~P.~Neupane,
  Phys.\ Rev.\  D {\bf 66}, 024044 (2002)
  [arXiv:hep-th/0202140];
  I.~P.~Neupane,
  Phys.\ Rev.\  D {\bf 69} (2004) 084011
  [arXiv:hep-th/0302132];
  \\  
  I.~P.~Neupane,
  Phys.\ Rev.\  D {\bf 67} (2003) 061501
  [arXiv:hep-th/0212092].
\bibitem{Dutta:2006vs}
  S.~Dutta and R.~Gopakumar,
  Phys.\ Rev.\  D {\bf 74} (2006) 044007
  [arXiv:hep-th/0604070].
\bibitem{Clunan:2004tb}
  T.~Clunan, S.~F.~Ross and D.~J.~Smith,
  Class.\ Quant.\ Grav.\  {\bf 21} (2004) 3447
  [arXiv:gr-qc/0402044].
\bibitem{Padilla:2003qi}
  A.~Padilla,
  Class.\ Quant.\ Grav.\  {\bf 20} (2003) 3129
  [arXiv:gr-qc/0303082].
\bibitem{Wald:1993nt}
  R.~M.~Wald,
  Phys.\ Rev.\  D {\bf 48} (1993) 3427
  [arXiv:gr-qc/9307038].
    
\bibitem{Kim:2007iw}
  H.~C.~Kim and R.~G.~Cai,
  Phys.\ Rev.\  D {\bf 77} (2008) 024045
  [arXiv:0711.0885 [hep-th]].
\bibitem{Dehghani:2002wn}
  M.~H.~Dehghani,
  Phys.\ Rev.\  D {\bf 67} (2003) 064017
  [arXiv:hep-th/0211191];
 \\
  M.~H.~Dehghani,
  Phys.\ Rev.\  D {\bf 69} (2004) 064024
  [arXiv:hep-th/0312030];
  \\
  M.~H.~Dehghani and R.~B.~Mann,
  Phys.\ Rev.\  D {\bf 73} (2006) 104003
  [arXiv:hep-th/0602243];
  \\
  M.~H.~Dehghani, G.~H.~Bordbar and M.~Shamirzaie,
  Phys.\ Rev.\  D {\bf 74} (2006) 064023
  [arXiv:hep-th/0607067].
\bibitem{Kunz:2006eh}
  J.~Kunz, F.~Navarro-Lerida and J.~Viebahn,
  Phys.\ Lett.\ B {\bf 639} (2006) 362
  [arXiv:hep-th/0605075].
\bibitem{Gibbons:2004js}
  G.~W.~Gibbons, H.~Lu, D.~N.~Page and C.~N.~Pope,
  Phys.\ Rev.\ Lett.\  {\bf 93}, (2004) 171102.   
 \bibitem{Deruelle:2004bj}
  N.~Deruelle and Y.~Morisawa,
  Class.\ Quant.\ Grav.\  {\bf 22} (2005) 933
  [arXiv:gr-qc/0411135].
\bibitem{tekin}
  S.~Deser, I.~Kanik and B.~Tekin,
  Class.\ Quant.\ Grav.\  {\bf 22} (2005) 3383
  [arXiv:gr-qc/0506057].

\bibitem{bs} 
G. T. Horowitz and A. Strominger, Nucl. Phys. B {\bf 360} (1991), 197.
 \bibitem{Chamblin:1999by}
  A.~Chamblin, S.~W.~Hawking and H.~S.~Reall,
  Phys.\ Rev.\  D {\bf 61} (2000) 065007
  [arXiv:hep-th/9909205].
\bibitem{Kastor:2006vw}
  D.~Kastor and R.~B.~Mann,
  JHEP {\bf 0604} (2006) 048
  [arXiv:hep-th/0603168].
\bibitem{Chamseddine:1999xk}
  A.~H.~Chamseddine and W.~A.~Sabra,
  Phys.\ Lett.\ B {\bf 477}, 329 (2000)
  [arXiv:hep-th/9911195];
  \\
  D.~Klemm and W.~A.~Sabra,
  Phys.\ Rev.\ D {\bf 62}, 024003 (2000)
  [arXiv:hep-th/0001131];
  \\
  W.~A.~Sabra,
  Phys.\ Lett.\ B {\bf 545}, 175 (2002)
  [arXiv:hep-th/0207128].
\bibitem{Brihaye:2007vm}
  Y.~Brihaye, E.~Radu and C.~Stelea,
  Class.\ Quant.\ Grav.\  {\bf 24} (2007) 4839
  [arXiv:hep-th/0703046].
 \bibitem{Brihaye:2007jua}
  Y.~Brihaye and E.~Radu,
  Phys.\ Lett.\  B {\bf 658} (2008) 164
  [arXiv:0706.4378 [hep-th]].
\bibitem{Bernamonti:2007bu}
  A.~Bernamonti, M.~M.~Caldarelli, D.~Klemm, R.~Olea, C.~Sieg and E.~Zorzan,
  JHEP {\bf 0801} (2008) 061
  [arXiv:0708.2402 [hep-th]].
  
  
\bibitem{Gregory:1993vy}
  R.~Gregory and R.~Laflamme,
  Phys.\ Rev.\ Lett.\  {\bf 70} (1993) 2837
  [arXiv:hep-th/9301052].
  
\bibitem{Brihaye:2007ju}
  Y.~Brihaye, T.~Delsate and E.~Radu,
  Phys.\ Lett.\  B {\bf 662} (2008) 264
  [arXiv:0710.4034 [hep-th]].

\bibitem{Delsate:2008kw}
  T.~Delsate,
  arXiv:0802.1392 [hep-th].
\bibitem{Kobayashi:2004hq}
  T.~Kobayashi and T.~Tanaka,
  Phys.\ Rev.\  D {\bf 71} (2005) 084005
  [arXiv:gr-qc/0412139].
  

\bibitem{Skenderis:2000in}
K.~Skenderis,
Int.\ J.\ Mod.\ Phys.\ A {\bf 16} (2001) 740,
[arXiv:hep-th/0010138];
\\
  M.~Henningson and K.~Skenderis,
  JHEP {\bf 9807} (1998) 023
  [arXiv:hep-th/9806087];
  \\
  M.~Henningson and K.~Skenderis,
  Fortsch.\ Phys.\  {\bf 48}, 125 (2000)
  [arXiv:hep-th/9812032].

\bibitem{Townsend:2001rg}
  P.~K.~Townsend and M.~Zamaklar,
  Class.\ Quant.\ Grav.\  {\bf 18} (2001) 5269
  [arXiv:hep-th/0107228].
\bibitem{Kudoh:2004hs}
  H.~Kudoh and T.~Wiseman,
  Phys.\ Rev.\ Lett.\  {\bf 94} (2005) 161102
  [arXiv:hep-th/0409111];
  \\
  B.~Kleihaus, J.~Kunz and E.~Radu,
  JHEP {\bf 0606} (2006) 016
  [arXiv:hep-th/0603119].
   
  \bibitem{Bastianelli:2000hi}
  F.~Bastianelli, S.~Frolov and A.~A.~Tseytlin,
  JHEP {\bf 0002}, 013 (2000)
  [arXiv:hep-th/0001041].
\bibitem{COLSYS}
 U. Ascher, J. Christiansen, R.~D. Russell,
 Mathematics of Computation {\bf 33} (1979) 659;
 ACM Transactions {\bf 7} (1981) 209.  
\bibitem{Dehghani:2005zm}
  M.~H.~Dehghani and R.~B.~Mann,
  Phys.\ Rev.\  D {\bf 72} (2005) 124006
  [arXiv:hep-th/0510083];
  \\
  M.~H.~Dehghani and S.~H.~Hendi,
  Phys.\ Rev.\  D {\bf 73} (2006) 084021
  [arXiv:hep-th/0602069].
\bibitem{Astefanesei:2004ji}
  D.~Astefanesei, R.~B.~Mann and E.~Radu,
  Phys.\ Lett.\  B {\bf 620} (2005) 1
  [arXiv:hep-th/0406050];
  \\
  D.~Astefanesei, R.~B.~Mann and E.~Radu,
  JHEP {\bf 0501} (2005) 049
  [arXiv:hep-th/0407110].
 
\bibitem{Kraus:1999di}
  P.~Kraus, F.~Larsen and R.~Siebelink,
  Nucl.\ Phys.\  B {\bf 563} (1999) 259
  [arXiv:hep-th/9906127].
\bibitem{Gubser:2000ec}
  S.~S.~Gubser and I.~Mitra,
  arXiv:hep-th/0009126;
\\
  S.~S.~Gubser and I.~Mitra,
  JHEP {\bf 0108} (2001) 018
  [arXiv:hep-th/0011127].
\bibitem{miki}
 D. Astefanesei,  N. Banarjee and S. Dutta,
{\it (Un)attractor black holes in higher derivative AdS gravity},
  [arXiv:0806.1334].
\bibitem{Radu:2005mj}
  E.~Radu and D.~H.~Tchrakian,
  Phys.\ Rev.\  D {\bf 73} (2006) 024006
  [arXiv:gr-qc/0508033].
  
  
  
\end{thebibliography}
\end{document}